\documentclass[journal]{IEEEtran}

\usepackage{epsf}
\usepackage{amsmath,amssymb}
\usepackage{graphicx}
\usepackage{color}
\usepackage{cite}
\usepackage{multirow,tabularx}
\usepackage{ifthen}
\usepackage{times}
\usepackage{stackrel}
\usepackage[left,pagewise]{lineno}

\newcommand{\hide}[1]{\ifthenelse{\boolean{false}}{#1}{}}

\newtheorem{theorem}{{\bf Theorem}}

\newtheorem{lemma}{{\bf Lemma}}

\newtheorem{corollary}{{\bf Corollary}}
\newtheorem{defn}{{\bf Definition}}

\newcommand{\qed}{\nobreak \ifvmode \relax \else
      \ifdim\lastskip<1.5em \hskip-\lastskip
      \hskip1.5em plus0em minus0.5em \fi \nobreak
      \vrule height0.75em width0.5em depth0.25em\fi}

\newcommand{\barr}{\begin{array}}
\newcommand{\earr}{\end{array}}

\newcommand{\mtx}[1]{{\bf #1}} 

\newcommand{\vect}[1]{\boldsymbol{#1}}

\newcommand{\expect}[1]{{\mathbb E}\left[{#1}\right]}

\newcommand{\bsp}{\begin{slide*}}
\newcommand{\esp}{\end{slide*}}
\newcommand{\bsl}{\begin{slide}}
\newcommand{\esl}{\end{slide}}

\newcommand{\var}[1]{{\mathrm{var}}\left[{#1}\right]}

\newcommand{\sth}{\text{th}}

\IEEEoverridecommandlockouts

\begin{document}
\title{On the Feasibility of Wireless Energy Transfer Using  Massive Antenna Arrays}
\author{Salil Kashyap, {\it Member, IEEE}, Emil Bj\"{o}rnson, {\it Member, IEEE}, and Erik G. Larsson, {\it Fellow, IEEE}
%
\thanks{This work was supported by the Swedish Research Council (VR) and ELLIIT. The authors are with the Division of Communication Systems, Dept.\ of Electrical Eng. (ISY) at Link\"{o}ping University, Link\"{o}ping, Sweden.}
\thanks{Emails: salil.kashyap@liu.se, emil.bjornson@liu.se, erik.g.larsson@liu.se}
\thanks{Parts of the results in this paper were presented at IEEE WPTC 2015, USA~\cite{Kashyap_2015_WPTC} and at IEEE SPAWC 2015, Sweden~\cite{Kashyap_2015_SPAWC}.}}

\IEEEaftertitletext{\vspace{-0.25\baselineskip}}

\maketitle
\setcounter{page}{1}

\begin{abstract}
We illustrate potential benefits of using massive antenna arrays for wireless energy transfer (WET). Specifically, we analyze probability of outage in WET over fading channels when a base station (BS) with multiple antennas beamforms energy to a wireless sensor node (WSN). Our analytical results show that by using massive antenna arrays, the range of WET can be increased for a given target outage probability. We prove that by using multiple-antenna arrays at the BS, a lower downlink energy is required to get the same outage performance, resulting into savings of radiated energy. We show that for energy levels used in WET, the outage performance with least-squares or minimum mean-square error channel estimates is same as that obtained based on perfect channel estimates. We observe that a strong line-of-sight component between the BS and WSN lowers outage probability. Furthermore, by deploying more antennas at the BS, a larger energy can be transferred reliably to the WSN at a given target outage performance for the sensor to be able to perform its main tasks. In our numerical examples, the RF power received at the input of the sensor is assumed to be on the order of a mW, such that the rectenna operates at an efficiency in the order of 50\%.
\end{abstract}

\begin{keywords}
Wireless energy transfer, massive MIMO, beamforming, outage probability, array gain
\end{keywords}

\IEEEpeerreviewmaketitle

\section{Introduction}
\label{sec:Introduction}
Wireless energy transfer (WET) is a promising energy harvesting technology where the destination node harvests energy from electromagnetic radiations instead of traditional wired energy sources~\cite{Ulukus_2015_ieeeJsac}. The use of WET can help increase the battery-lifetime of energy-constrained wireless sensor nodes (WSNs) that are used for applications such as intelligent transportation, intrusion detection, and aircraft structural monitoring~\cite{Popovic_2013_procIEEE}. Furthermore, WET can be used to charge low power devices such as temperature and humidity meters and liquid crystal displays~\cite{Ding_2015_ieeeCommMag}. Even low-end computation, sensing, and communication can be performed by harvesting energy from ambient radio frequency (RF) signals including TV, cellular networks, and Wi-Fi transmissions~\cite{Kellogg_2014_SIGCOMM}.

However, there are several challenges that must be addressed in order to implement WET. Firstly, only a small fraction of the energy radiated by an energy transmitter can be harvested by the WSN which severely limits the range of WET~\cite{Popovic_2013_procIEEE, Visser_2013_ieee}. Secondly, the received power levels that are suitable for wireless information transfer are not suitable for energy transfer, where the absolute received power is of interest and not the signal-to-noise ratio (SNR).

Massive multiple input multiple output (MIMO) systems, where the base station (BS) uses antenna arrays equipped with a few hundred antennas, have recently emerged as a leading $5$G wireless communications technology that offer orders of magnitude better data rates and energy efficiency than current wireless systems~\cite{Larsson_2014_ieeeCommMag}. Potentially, the use of massive arrays could significantly boost the performance of WET as well.

\subsection{Focus and Contributions}
\label{subsec:Focus and Contributions}
We consider a scenario where a multi-antenna BS communicates with and transfers RF power to a WSN.  The motivation of using an array of antennas is that the BS can exploit an array gain, resulting from coherent combination of the signals transmitted from each antenna, if it knows the channel response.  This array gain in turn may increase the operating range and/or decrease the amount of transmit energy needed to satisfy a given energy harvesting constraint. The drawback is that the wireless channel between the BS and the WSN fluctuates so that the channel state information (CSI) needs to be acquired on a regular basis to enable coherent combining.

\begin{figure}
\centering \includegraphics[width=.7\linewidth]{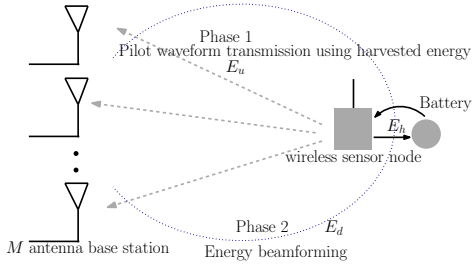}
\caption{Proposed two-phase protocol: Parameters are explained in Section~\ref{sec:System Model}.}
\label{fig:system_model_two_phases}
\vspace{-0.5cm}
\end{figure}

The communication between the array and the sensor takes place in two phases as shown in Figure~\ref{fig:system_model_two_phases}. In the first phase, the sensor utilizes energy stored in a battery or capacitor to transmit a pilot waveform which is measured at each antenna in the BS array, in order to estimate the channel impulse response from the sensor.  In the second phase, the array beamforms energy to the sensor, using the estimated channel responses and exploiting reciprocity\footnote{We consider time-division duplexing (TDD) mode of communication and both the uplink and the downlink communication take place over the same narrowband channel. We adopt the widely used reciprocity assumption, which implies that the channel gain from the BS to a WSN is the same as the channel gain from the WSN to the BS~\cite{Marzetta_2006_ieeeSigProc}. Most physical channels satisfy this assumption, but the transceiver hardware might not satisfy this condition unless calibration algorithms are applied~\cite{Zetterberg_2011_Eurasip}. However, there is substantial evidence that such calibration can be performed accurately and rather infrequently~\cite{Vieira_2014_globecom}.} of the propagation channel.  The energy harvested by the sensor is used to recharge its capacitor or battery, and needed in turn for pilot transmission in phase one of the next round and also to perform the main tasks of the sensor.  In addition, both phases may involve communication of information, although that is out of the scope of this paper.

The main questions asked and answered in this paper are: 1) What array gain can the massive MIMO setup provide, i.e., how does the required uplink pilot energy (and thereby the energy storage requirements at the sensor, and the required array transmit energy) scale with the number of antennas in the array taking into account that all channel responses are estimated from pilots? The goal is to determine the scaling laws for energy transfer using large arrays. 2) How does the number of antennas at the BS depend on the path loss or the distance between the BS and the WSN? 3) How do the answers to the previous questions depend on propagation conditions and the correlation between the adjacent antennas? 4) What role does power adaptation based on the estimated CSI play in improving the outage performance?

To this end, we derive new expressions that are valid for any generic path loss model for the probability of outage in energy transfer, defined here as the probability that the energy harvested by the WSN is less than the energy that it spends on uplink pilots plus the processing energy.\footnote{We refer to the energy that is needed by the sensor node to perform its main tasks as the processing energy.} We derive expressions for both perfect CSI and imperfect CSI based on least squares (LS) or minimum mean square error (MMSE) channel estimation and for both Rayleigh fading (without dominant channel components) and Rician fading (with dominant channel components). We consider not just the scenarios where the downlink array transmit energy is fixed but also those where it is adapted based on the channel conditions. We present numerical results to quantify the combined effects of path loss, energy spent on uplink pilot signaling, the downlink energy, the processing energy, the energy harvesting efficiency, the Rician $K$-factor, power adaptation, the correlation between adjacent antennas, and imperfect CSI on the probability of outage in energy transfer. To summarize, one of the main goals of this paper is to estimate the link budget in order to determine the feasibility of a system that performs WET using multi-antenna arrays. We next discuss the relevant literature on WET using multi-antenna arrays.

\subsection{Related Literature}
\label{subsec:Related Literature}
The optimal uplink pilot power and the number of antennas at the sensor that need to be trained so as to maximize the net average harvested energy at the sensor node was characterized in~\cite{Zeng_2015_ieeeComm}. However, reference~\cite{Zeng_2015_ieeeComm} did not consider the possibility of an outage in energy transfer. The amount of time that must be allocated for channel estimation and for WET in order to maximize the harvested energy for a multiple input single output (MISO) system was investigated in~\cite{Gang_2014_ieeeSigProc}. In~\cite{Liu_2014_ieeeComm}, a wireless powered communication network with one multi-antenna BS and a set of single antenna users was studied for joint downlink (DL) energy transfer and uplink (UL) information transmission via spatial division multiple access. The aim was to maximize the minimum data throughput among all users by optimizing the DL-UL time allocation, DL energy beamforming, and UL transmit power allocation. While, the optimal training design to maximize the net average harvested energy at the sensor over frequency-selective channels was studied in~\cite{Zeng_2015_ieeeComm_freq}, energy transfer in the downlink to maximize the minimum rate among all users was studied in~\cite{Gang_2015_ieeeJsac}.

Simultaneous wireless information and power transfer (SWIPT), where a multi-antenna BS sends information and energy simultaneously to several users which then perform information decoding or energy harvesting was studied in~\cite{Xu_2014_ieeeSigProc, Chen_2015_ieeeCommMag, Zhang_2013_ieeeTWC}. The authors in~\cite{Liu_2013_ieeeTWC} investigated when the receiver should switch from the information decoding mode to the energy harvesting mode based on the instantaneous channel and interference conditions so as to achieve various trade-offs between wireless information transfer and energy harvesting. Receiver design for SWIPT over a point-to-point wireless link was investigated in~\cite{Zhou_2013_ieeeComm}. In~\cite{Huang_2014_ieeeTWC}, the authors studied a hybrid network architecture that overlays an uplink cellular network with randomly deployed power beacons for charging the mobile devices wirelessly. The tradeoffs between the network parameters such as transmission powers and the densities of BSs and power beacons were derived under an outage constraint on the data links. Using a stochastic geometry approach, upper bounds on both transmission and power outage probabilities for a downlink SWIPT system with ambient RF transmitters was developed in~\cite{Lu_2015_wcnc}. Energy transfer in relay systems to simultaneously harvest energy and process information was investigated in~\cite{Nasir_2013_ieeeTWC, Ding_2014_ieeeTWC, Zeng_2015_WicommLet}.

In contrast to most existing works in the literature, we focus on ascertaining whether the use of large antenna arrays could substantially extend the feasible range of WET while maintaining the receive power level in the same order such that a reasonable rectenna efficiency can be maintained.

The paper is organized as follows:
We present the system model in Section~\ref{sec:System Model}. The analysis of the
probability of outage in energy transfer for different scenarios is given
in Section~\ref{sec: Analysis of Probability of Outage in Energy Transfer}
and summarized in Tables~\ref{table:outage_probability_without_adaptation} and~\ref{table:outage_probability_with_adaptation}.
Numerical results and our conclusions follow in Section~\ref{sec:Numerical Results} and
Section~\ref{sec: Conclusions}, respectively.

The notation $X \sim {\cal CN}(0,\delta)$ means that $X$ is a circularly symmetric complex
Gaussian RV with zero mean and variance $\delta$, and $\vect{x} \sim
{\cal CN}(\vect{m},\mtx{C})$ means that $\vect{x}-\vect{m}$ is a circularly symmetric complex
Gaussian random vector with covariance matrix $\mtx{C}$ and zero mean vector. The expectation of a RV $X$ is denoted by $\mathbb{E}\left[X\right]$. The probability density function (PDF) of a RV $X$ is denoted by $f_{X}(x)$. The notation $(\cdot)^{\dag}$ denotes conjugate transpose. Given a complex number $z$, we denote its real part by $\mathrm{Re}(z)$ and imaginary part by $\mathrm{Im}(z)$.

\section{System Model}
\label{sec:System Model}

We consider a block-fading channel model in which the channel impulse response from each antenna at the BS to the WSN remains constant during a coherence interval of $\tau$ seconds. The channel realizations are random and they are independent across blocks. We, therefore, need to estimate the channel after every coherence interval. We assume TDD mode of communication so that the channel from the BS to the WSN referred to as the downlink channel is the same as the channel from the WSN to the BS referred to as the uplink channel. Therefore, the BS can take advantage of channel reciprocity and make channel measurements using uplink signals.

We focus on a wireless network where a BS with $M$ antennas is used to transfer RF energy to a single antenna WSN that has energy harvesting capabilities. We consider a scenario where a line-of-sight (LoS) link might be present between the BS and the WSN and for which the complex channel gain vector $\vect{h}$ from the BS to the WSN can be represented by the Rician fading model as~\cite{Farrokhi_2001_commLet}
\begin{equation}
\vect{h} = \sqrt{\frac{\beta K}{K+1}} \vect{h}_d + \sqrt{\frac{\beta}{K+1}} \vect{h}_s,
\label{LOS_model}
\end{equation}
where $\vect{h}_d \in \mathbb{C}^{M\rm{x}1}$ is a deterministic vector containing the line-of-sight and the specular components of the channel, $\beta$ denotes distance-dependent path loss, $K$ is the Rician factor defined as the ratio of the deterministic to the scattered power, and $\vect{h}_s \in \mathbb{C}^{M\rm{x}1}$ denotes the scattered components of the channel and is a random vector with i.i.d. zero mean unit variance circular symmetric complex Gaussian entries. Furthermore,
$\vect{h}_d = \left[\sqrt{\alpha_0}~~~\sqrt{\alpha_1} e^{j \theta_1 (\phi)}~~~ \cdots ~~~\sqrt{\alpha_{M-1}} e^{j \theta_{(M-1)}(\phi)}\right]^{T}$
where $\alpha_i$, $i = 0,\ldots, M-1$ denotes the gain of the $i^{\sth}$ antenna which takes a large value if the $i^{\sth}$ link is good and a small value if it is bad, $\theta_i(\phi)$, $i = 1,\ldots, M-1$ is the phase shift of the $i^{\text{th}}$ antenna with respect to the reference antenna and $\phi$ is the angle of departure/arrival of the specular component.
Thus, $\vect{h} \sim \mathcal{CN}\left(\vect{\mu}, \mtx{\Lambda}_h\right)$, where $\vect{\mu} = \sqrt{\frac{\beta K}{K+1}} \left[\sqrt{\alpha_0}~~~\sqrt{\alpha_1} e^{j \theta_1 (\phi)}~~~ \cdots ~~~\sqrt{\alpha_{M-1}} e^{j \theta_{(M-1)}(\phi)}\right]^{T}$ and $\mtx{\Lambda}_h = \frac{\beta}{K+1} \mathbf{I}_M$.
By varying $K$, the model discussed above captures a general class of wireless channels spanning from a rich-scattering Rayleigh fading channel ($K=0$) to a completely deterministic channel ($K \rightarrow \infty$).

\newcommand{\expecth}[1]{{\mathbb{E}_{\mathbf{h}}}\left[{#1}\right]}
\newcommand{\expecty}[1]{{\mathbb{E}_{\mathbf{y}}}\left[{#1}\right]}
\subsection{Uplink Pilot Signaling and Channel Estimation}
\label{subsec:Pilot Signaling and Channel Estimation}
The signal\footnote{This is the complex baseband representation of a physical quantity that is proportional to the voltage measured across the load connected to the BS antenna. The proportionality constant in turn depends on the load resistor used.} $\vect{y}(t)$ received at the BS when the WSN transmits a continuous-time pilot signal $\sqrt{E_u} p(t)$ of duration $T < \tau$ such that $\int_{0}^{T} |p(t)|^2 \, dt  =1$, is given by
\begin{equation}
\vect{y}(t) = \sqrt{E_u} \vect{h} p(t) + \vect{w}(t),~~\text{for}~t \in \left[0, T\right],
\end{equation}
where $E_u$ is the uplink pilot energy in Joule, and $\vect{h} \in \mathbb{C}^{M\rm{x}1}$ is the channel gain vector from the WSN to the $M$ antennas at the BS as defined in~\eqref{LOS_model}. Also, $\vect{w}(t)$ is the thermal noise vector at the BS that is independent of $\vect{h}$. The objective of the pilot signaling is to estimate $\vect{h}$ given $\vect{y}(t)$.

Now, a sufficient statistic for estimating $\vect{h}$ at the BS is
\begin{equation}
\vect{y} = \int_{0}^{T} p^{\ast}(t) \vect{y}(t) dt = \sqrt{E_u} \vect{h} + \vect{w},
\end{equation}
where $\vect{w} \in \mathbb{C}^{M\rm{x}1}$ is the circular symmetric complex additive white Gaussian noise (AWGN) at the BS. Furthermore, $\vect{w} \sim \mathcal{CN}(\vect{0}, N_0 \mathbf{I}_M)$, where $N_0$ is the noise power spectral density in Joule.
There are different ways of estimating $\vect{h}$ depending on which type of a priori information that is available at the BS.
\newcommand{\LS}{\text{LS}}
\newcommand{\MMSE}{\text{MMSE}}

\subsubsection{LS Channel Estimation}
\label{subsubsec:Least Squares Channel Estimation}
This can be used when the distributions of the noise and the channel are not known a priori. The LS channel estimate is also the maximum likelihood estimate in an AWGN setting. Thus, given the observation vector $\vect{y}$ at the BS, the LS channel estimate $\widehat{\vect{h}}_{\LS}$ of $\vect{h}$ is given by~\cite{kay_book_vol1}
\begin{equation}
\widehat{\vect{h}}_{\LS} = \frac{\vect{y}}{\sqrt{E_u}}.
\label{LS_estimate_LoS_Model}
\end{equation}
This can be simplified to obtain
\begin{equation}
\widehat{\vect{h}}_{\LS} = \vect{h} + \widetilde{\vect{h}}_{\LS},
\end{equation}
where $\widetilde{\vect{h}}_{\LS} \sim \mathcal{CN}(\vect{0}, \frac{N_0}{E_u} \mathbf{I}_M)$ is the estimation error~\cite{kay_book_vol1}, which is circularly symmetric and is independent of  $\vect{h}$, since it is a linear function of $\vect{w}$. Furthermore, $\widehat{\vect{h}}_{\LS} \sim \mathcal{CN} \left(\vect{\mu}, \frac{\beta E_u + (K+1) N_0}{E_u (K+1)} \mathbf{I}_M\right)$.

\subsubsection{MMSE Channel Estimation}
\label{subsubsec:Minimum Mean Square Error Channel Estimation}
If the distribution of the channel and noise are known a priori, MMSE channel estimation can be used. In that case, the MMSE estimate $\widehat{\vect{h}}_{\MMSE}$ of $\vect{h}$ is~\cite{kay_book_vol1}
\begin{equation}
\widehat{\vect{h}}_{\MMSE} = \expect{\vect{h}|\vect{y}} = \expect{\vect{h}} + \mathrm{cov}\left(\vect{h}, \vect{y}\right) \left(\mathrm{cov}\left(\vect{y}, \vect{y}\right)\right)^{-1} (\vect{y} - \expect{\vect{y}}),
\label{MMSE_estimate_LoS_Model}
\end{equation}
where $\mathrm{cov}\left(\vect{h}, \vect{y}\right)$ is the cross-covariance matrix of $\vect{h}$ and $\vect{y}$ and $\mathrm{cov}\left(\vect{y}, \vect{y}\right)$ is the covariance matrix of $\vect{y}$. It is straightforward to show that $\mathrm{cov}\left(\vect{h}, \vect{y}\right)  = \frac{\beta\sqrt{E_u}}{K+1} \mathbf{I}_M$ and $\mathrm{cov}\left(\vect{y}, \vect{y}\right)  = \frac{\beta E_u + (K+1) N_0}{K+1} \mathbf{I}_M$.
Therefore, the MMSE estimate of $\vect{h}$ in~\eqref{MMSE_estimate_LoS_Model} can be simplified to obtain
\begin{equation}
\widehat{\vect{h}}_{\MMSE} = \vect{h} + \widetilde{\vect{h}}_{\MMSE},
\end{equation}
where $\widetilde{\vect{h}}_{\MMSE} \sim \mathcal{CN}(\vect{0}, \frac{\beta N_0}{\beta E_u + (K+1) N_0} \mathbf{I}_M)$ is the estimation error~\cite{kay_book_vol1}, which is circularly symmetric since it consists of noise at the BS and a part of the true channel, both of which are circularly symmetric. Also, it is independent of  $\widehat{\vect{h}}_{\MMSE}$. Furthermore, $\widehat{\vect{h}}_{\MMSE} \sim \mathcal{CN}\left(\vect{\mu}, \frac{\beta^2 E_u}{(\beta E_u + (K+1) N_0)(K+1)} \mathbf{I}_M\right)$.


\subsection{Transmit Beamforming Based on the Estimated Channel}
\label{subsec:Transmit Beamforming Based on the Estimated Channel}
In this subsection, we will see how the BS performs transmit beamforming based on either the LS or the MMSE channel estimate and also characterize the energy harvested.
\subsubsection{Transmit Beamforming Based on the LS Channel Estimate}
\label{subsubsec:Transmit Beamforming Based on the LS Channel Estimate}
Given the channel estimate $\widehat{\vect{h}}_{\LS}$, the BS performs transmit beamforming of energy: it selects the signals emitted from the different antennas so that they add up coherently at the WSN, i.e., maximizes the harvested received energy. Thus, on the downlink, it transmits $\vect{x}(t) = \sqrt{E_d} \frac{\widehat{\vect{h}}_{\LS}^{\dag}}{||\widehat{\vect{h}}_{\LS}||}p^{\prime}(t)$, where $E_d$ is the downlink array transmit energy in Joule, and $p^{\prime}(t)$ is a unit energy continuous-time pulse of duration $T^{\prime}$. Also, $T+T^{\prime} \leq \tau$. The continuous-time signal $y^{\prime}(t)$ received by the WSN is
\begin{equation}
y^{\prime}(t) = \sqrt{E_d} \frac{\widehat{\vect{h}}_{\LS}^{\dag}\vect{h}}{||\widehat{\vect{h}}_{\LS}||}p^{\prime}(t) + w^{\prime}(t),~~\text{for}~t \in \left[0, T^{\prime}\right],
\end{equation}
where $w^{\prime}(t)$ is the thermal noise at the WSN. Let $\eta$ denote the energy harvesting efficiency of the WSN. Then, the energy harvested $E_h$ in Joule is
\begin{equation}
E_h = \eta E_d \left|\frac{\widehat{\vect{h}}_{\LS}^{\dag}\vect{h}}{||\widehat{\vect{h}}_{\LS}||}\right|^{2}.
\label{eq:energy_harvested}
\end{equation}
Note that $E_h$ is a random variable since both $\vect{h}$ and $\widehat{\vect{h}}_{\LS}$ are random. We have neglected the contribution from $w^{\prime}(t)$ to $E_h$, since it is negligible.

Let us now define
\begin{equation}
\label{eq:psi_ls}
\Psi_{\LS} \triangleq \frac{\widehat{\vect{h}}_{\LS}^{\dag}\vect{h}}{||\widehat{\vect{h}}_{\LS}||},
\end{equation}
which is the RV in the harvested energy expression in~\eqref{eq:energy_harvested}.
We state below a result that will be used in the performance analysis in Section~\ref{subsec: Analysis with LS Channel Estimate}.
\begin{lemma}
\label{Lem_LS}
Given the LS channel estimate $\widehat{\vect{h}}_{\LS}$, $\Psi_{\LS}$ is a complex Gaussian RV with conditional mean
\begin{multline}
\expect{\Psi_{\LS}|\widehat{\vect{h}}_{\LS}} = \frac{\beta E_u}{\beta E_u + (K+1) N_0}||\widehat{\vect{h}}_{\LS}||  \\ + \frac{(K+1) N_0}{\beta E_u + (K+1) N_0} \frac{\widehat{\vect{h}}_{\LS}^{\dag}\vect{\mu}}{||\widehat{\vect{h}}_{\LS}||},
\label{cond_mean_LS}
 \end{multline}
 and conditional variance
 \begin{equation}
 \var{\Psi_{\LS}|\widehat{\vect{h}}_{\LS}}  = \frac{\beta N_0}{\beta E_u + (K+1) N_0}.
\label{cond_var_LS}
 \end{equation}
\end{lemma}
\begin{IEEEproof}
The proof is given in Appendix~\ref{app_conditional_stats_Psi_LSestimate}.
\end{IEEEproof}
\begin{corollary}
For Rayleigh fading ($K = 0$), the RV $\Psi_{\LS}$ given $\widehat{\vect{h}}_{\LS}$ is distributed as
\begin{equation}
\Psi_{\LS}|\widehat{\vect{h}}_{\LS} \sim \mathcal{CN}\left(\frac{\beta E_u}{\beta E_u + N_0}||\widehat{\vect{h}}_{\LS}|| , \frac{\beta N_0}{\beta E_u + N_0}\right).
\label{eq:statistics_of_psi_given_hhat_Rayleigh}
\end{equation}
\end{corollary}


\subsubsection{Transmit Beamforming Based on the MMSE Channel Estimate}
\label{subsubsec:Transmit Beamforming Based on the MMSE Channel Estimate}
 If the BS performs transmit beamforming given the MMSE estimate $\widehat{\vect{h}}_{\MMSE}$ and on the downlink transmits $\vect{x}(t) = \sqrt{E_d} \frac{\widehat{\vect{h}}_{\MMSE}^{\dag}}{||\widehat{\vect{h}}_{\MMSE}||}p^{\prime}(t)$ instead, then
the signal $y^{\prime}(t)$ received by the WSN on the downlink is
\begin{equation}
y^{\prime}(t) = \sqrt{E_d} \frac{\widehat{\vect{h}}_{\MMSE}^{\dag}\vect{h}}{||\widehat{\vect{h}}_{\MMSE}||}p^{\prime}(t) + w^{\prime}(t).
\end{equation}

When using the MMSE estimate for beamforming, the energy harvested $E_h$ in Joule is
\begin{equation}
E_h = \eta E_d \left|\frac{\widehat{\vect{h}}_{\MMSE}^{\dag}\vect{h}}{||\widehat{\vect{h}}_{\MMSE}||}\right|^{2}.
\label{eq:energy_harvested_MMSE}
\end{equation}
 We next characterize this RV that is based on the MMSE estimate.
To that end, let us define
\begin{equation}
\Psi_{\MMSE} \triangleq \frac{\widehat{\vect{h}}_{\MMSE}^{\dag}\vect{h}}{||\widehat{\vect{h}}_{\MMSE}||}.
\end{equation}
We state below a result that will be used in the performance analysis in Section~\ref{subsec: Analysis with MMSE Channel Estimate}.
\begin{lemma}
\label{Lem_MMSE}
Given the MMSE channel estimate $\widehat{\vect{h}}_{\MMSE}$, $\Psi_{\MMSE}$ is a complex Gaussian RV with
conditional mean
\begin{equation}
 \expect{\Psi_{\MMSE}|\widehat{\vect{h}}_{\MMSE}} = ||\widehat{\vect{h}}_{\MMSE}||
\label{eq:cond_mean}
\end{equation}
and conditional variance
\begin{equation}
\var{\Psi_{\MMSE}|\widehat{\vect{h}}_{\MMSE}} = \frac{\beta N_0}{\beta E_u + (K+1) N_0}.
 \label{eq:cond_var}
\end{equation}
\end{lemma}
\begin{IEEEproof}
The proof is given in Appendix~\ref{app_conditional_stats_Psi_MMSEestimate}.
\end{IEEEproof}

\begin{corollary}
For Rayleigh fading ($K = 0$), the RV $\Psi_{\MMSE}$ given $\widehat{\vect{h}}_{\MMSE}$ is distributed as
\begin{equation}
\Psi_{\MMSE}|\widehat{\vect{h}}_{\MMSE} \sim \mathcal{CN}\left(||\widehat{\vect{h}}_{\MMSE}|| , \frac{\beta N_0}{\beta E_u + N_0}\right).
\label{eq:statistics_of_psi_given_hhat_Rayleigh_MMSE}
\end{equation}
\end{corollary}

The conditional statistics derived in this section are used subsequently in the analysis of the probability of outage in energy transfer in the next section.
\newcommand{\hhat}{\widehat{\vect{h}}}
\newcommand{\psit}{\widetilde{\Psi}}

\section{Analysis of Probability of Outage in Energy Transfer}
\label{sec: Analysis of Probability of Outage in Energy Transfer}
Ideally, we want the energy harvested $E_h$ to be greater than the sum of the energy $E_u$ spent on uplink pilot signaling and the processing energy $E_p$ that is required by the sensor to perform its main tasks. However, this cannot always be guaranteed on fading channels. In this section, we compute the probability of outage in energy transfer.
\begin{defn}
The probability of outage in energy transfer $P_o$ is defined mathematically as
\begin{equation}
P_o = \Pr (E_h \leq E_u + E_p).
\end{equation}
\end{defn}

We compute this probability of outage for scenarios when the BS has an LS or an MMSE estimate of the channel from itself to the WSN.
As a baseline, we also consider the case of perfect CSI, in which case the BS knows $\vect{h}$ exactly. This reference case gives us a bound in terms of the best outage performance  that can be achieved and we include it to understand when the uplink pilot is the limiting factor. We develop outage expressions not only for scenarios when the downlink array transmit energy $E_d$ is fixed but also when $E_d$ is adapted based on the channel conditions.
Results for different scenarios are summarized in Tables~\ref{table:outage_probability_without_adaptation} and~\ref{table:outage_probability_with_adaptation}.

\begin{table*}[t]
\caption{Probability of outage in energy transfer for different scenarios without power adaptation}
\centering
\label{table:outage_probability_without_adaptation}
\begin{tabular}{l l}
  \hline
   Scenario & Probability of outage in energy transfer \\
  \hline
  Perfect CSI, $K = 0$ &$\frac{\gamma\left(M, \frac{E_u+E_p}{\eta \beta E_d}\right)}{(M-1)!}$  \\
  \hline
  LS estimation, $K = 0$ & $1-\frac{\beta E_u}{\beta E_u + N_o} \exp\left(-\frac{E_u+E_p}{\eta \beta E_d}\right) \sum_{k=0}^{M-1} \epsilon_{k} \left(\frac{N_o}{\beta E_u + N_o}\right)^{k} L_{k}\left(-\frac{E_u(E_u+E_p)}{\eta E_d N_o}\right)$  \\
  \hline
  MMSE estimation, $K = 0$ & $1-\frac{\beta E_u}{\beta E_u + N_o} \exp\left(-\frac{E_u+E_p}{\eta \beta E_d}\right) \sum_{k=0}^{M-1} \epsilon_{k} \left(\frac{N_o}{\beta E_u + N_o}\right)^{k} L_{k}\left(-\frac{E_u(E_u+E_p)}{\eta E_d N_o}\right)$  \\
  \hline
  Perfect CSI, $K \neq 0$ &$1-Q_{M}\left(\sqrt{2 K \sum_{i=0}^{M-1}\alpha_i}, \sqrt{\frac{2(K+1) (E_u+E_p)}{\eta \beta E_d}}\right)$  \\
  \hline
  LS estimation, $K \neq 0$ & No closed-form, can be evaluated numerically using~\eqref{eq:outage_prob_imperfect_CSI_LS_K_nonzero}  \\
  \hline
  MMSE estimation, $K \neq 0$ & Single integral form, can be evaluated numerically using~\eqref{eq:exact_outage_expression_imperfect_CSI_MMSE_Knon_zero}  \\

  \hline
\end{tabular}
\end{table*}

\subsection{Analysis with Perfect CSI}
\label{subsec: Analysis with Perfect CSI}
As mentioned before, the channel estimation is considered error-free if we spend $E_u$ on uplink pilot signaling and there is no noise in the estimation process. In this subsection, we first investigate the scenario where $E_d$ is fixed. Thereafter, we analyze the probability of outage with power adaptation, where $E_d$ is varied based on the instantaneous channel conditions.
\subsubsection{Without Power Adaptation}
With fixed $E_d$, $P_o$ is given as follows:
\begin{theorem}
\label{th:outage_prob_perfect_CSI_K_nonzero}
For a Rician fading channel, the probability of outage $P_o$ in energy transfer with perfect CSI and with fixed $E_d$ is given by
\begin{equation}
P_o = 1-Q_{M}\left(\sqrt{2 K \sum_{i=0}^{M-1} \alpha_i}, \sqrt{\frac{2(K+1) (E_u+E_p)}{\eta \beta E_d}}\right),
\label{eq:outage_prob_perfect_CSI_K_nonzero}
\end{equation}
where $Q_{M}(\cdot,\cdot)$ is the $M^{\text{th}}$ order Marcum-Q function~\cite[Eqn.~(4.59)]{simon_alouini_book}.
\end{theorem}
\begin{IEEEproof}
The proof is given in Appendix~\ref{app_outage_prob_perfect_CSI_K_nonzero}.
\end{IEEEproof}
Note that, if $\alpha_i = 1$, for all $i = 0, \ldots, M-1$, then $P_o = 1-Q_{M}\left(\sqrt{2 K M}, \sqrt{\frac{2(K+1) (E_u+E_p)}{\eta \beta E_d}}\right)$.
Next, we state the probability of outage in energy transfer for a Rayleigh fading channel.
\begin{corollary}
\label{cor:outage_rayleigh_perfect_CSI}
For Rayleigh fading ($K = 0$) and with fixed $E_d$, $P_o$ is given as follows:
\begin{equation}
P_o  = 1-Q_{M}\left(0, \sqrt{\frac{2 (E_u +E_p)}{\eta \beta E_d}}\right) = \frac{\gamma\left(M, \frac{E_u+E_p}{\eta \beta E_d}\right)}{(M-1)!},
\label{eq:outage_prob_perfect_CSI_K_0}
\end{equation}
where the second equality in~\eqref{eq:outage_prob_perfect_CSI_K_0} follows from the identity in~\cite[Eqn~(4.71)]{simon_alouini_book} and $\gamma(\cdot,\cdot)$ is the lower incomplete Gamma function~\cite[Eqn.~(6.5.2)]{abramowitz_stegun}.
\end{corollary}

\subsubsection{With Power Adaptation}
The probability of outage in energy transfer when $E_d = \frac{\rho}{||\vect{h}||^2}$, where $\rho$ is the power control parameter, i.e., $E_d$ is adapted based on the channel conditions is given as follows:\footnote{Inverse channel inversion can be impractical for small number of antennas, but not as $M$ increases. This is another benefit of having an array of antennas.}
\begin{theorem}
\label{th:outage_prob_perfect_CSI_K_nonzero_adapt}
For a Rician or a Rayleigh fading channel, the probability of outage $P_o$ in energy transfer with perfect CSI and with power adaptation can be
made zero if and only if
\begin{equation}
\rho \geq \frac{E_u+E_p}{\eta}.
\end{equation}
\end{theorem}
\begin{IEEEproof}
The proof is given in Appendix~\ref{app_outage_prob_perfect_CSI_K_nonzero_adapt}.
\end{IEEEproof}

\subsection{Analysis with LS Channel Estimation}
\label{subsec: Analysis with LS Channel Estimate}
We now investigate the probability of outage in energy transfer when the BS performs transmit beamforming using the LS channel estimate, first with fixed $E_d$ and thereafter with $E_d$ adapted based on the estimated channel conditions.
\subsubsection{Without Power Adaptation}
With fixed $E_d$, $P_o$ for LS channel estimation is as follows:
\begin{theorem}
\label{th:outage_prob_imperfect_CSI_LS_K_nonzero}
For a Rician fading channel, the probability of outage $P_o$ in energy transfer with LS channel estimate and for a fixed $E_d$ is
\begin{equation}
P_o  \!=\! \expect{\!1\!-\!Q_{1}\!\left(\!\!\sqrt{\zeta (\widehat{\vect{h}}_{\LS})}\!,\! \sqrt{\!\frac{2(E_u\!+\!E_p)\!(\beta E_u \!+\! (K+1) N_0)}{\eta \beta E_d N_0}}\!\right)\!},
\label{eq:outage_prob_imperfect_CSI_LS_K_nonzero}
\end{equation}
where
\begin{multline}
\zeta (\widehat{\vect{h}}_{\LS}) =  \frac{2\left(\beta E_u ||\hhat_{\LS}||^{2} + \mathrm{Re}\left(\hhat_{\LS}^{\dag}\vect{\mu}\right) (K+1) N_0 \right)^2}{\beta N_0(\beta E_u + (K+1) N_0)||\hhat_{\LS}||^{2}} \\ \hspace{2cm} + \frac{2 N_0 (K+1)^{2}}{\beta (\beta E_u + (K+1) N_0)}\left(\frac{\mathrm{Im}\left(\widehat{\vect{h}}_{\LS}^{\dag}\vect{\mu}\right)}{||\widehat{\vect{h}}_{\LS}||}\right)^{2}.
\label{eq:findpdf}
\end{multline}
\end{theorem}
\begin{IEEEproof}
The proof is given in Appendix~\ref{app_outage_prob_imperfect_CSI_LS_K_nonzero}.
\end{IEEEproof}
To compute~\eqref{eq:outage_prob_imperfect_CSI_LS_K_nonzero} in closed-form, we need to find the distribution of $\zeta (\widehat{\vect{h}}_{\LS})$ given in~\eqref{eq:findpdf}. This is analytically intractable but the expectation in~\eqref{eq:outage_prob_imperfect_CSI_LS_K_nonzero} is easily evaluated numerically. A closed-form expression for the outage probability for a Rayleigh fading channel can, however, be obtained as stated below.
\begin{corollary}
\label{th:outage_prob_imperfect_CSI_K_0}
For a Rayleigh fading channel ($K = 0$), the probability of outage $P_o$ in energy transfer with LS channel estimate and fixed $E_d$ is
\begin{multline}
P_o  = 1-\frac{\beta E_u}{\beta E_u + N_o} \exp\left(-\frac{E_u+E_p}{\eta \beta E_d}\right)\\ \hspace{1cm}\times \sum_{k=0}^{M-1} \epsilon_{k} \left(\frac{N_o}{\beta E_u + N_o}\right)^{k} L_{k}\left(-\frac{E_u(E_u+E_p)}{\eta E_d N_o}\right),
\label{eq:exact_outage_expression_imperfect_CSI_K_0_LS}
\end{multline}
where $L_{k}(\cdot)$ is the $k^{\text{th}}$ Laguerre polynomial and
\begin{equation}
\epsilon_{k} = \begin{cases} 1, & k < M-1, \\
1+\frac{N_o}{\beta E_u},& k = M-1. \end{cases}
\label{eq:lag_coeff}
\end{equation}
\end{corollary}
\begin{IEEEproof}
The proof is given in Appendix~\ref{app_outage_prob_imperfect_CSI_K_0}.
\end{IEEEproof}

Based on Theorems~\ref{th:outage_prob_perfect_CSI_K_nonzero} and~\ref{th:outage_prob_imperfect_CSI_LS_K_nonzero} and Corollaries~\ref{cor:outage_rayleigh_perfect_CSI} and~\ref{th:outage_prob_imperfect_CSI_K_0}, we observe the following: For fixed $M$, $E_u$, $E_p$, $\eta$, and $\beta$, using~\eqref{eq:outage_prob_perfect_CSI_K_0} for perfect CSI or using~\eqref{eq:exact_outage_expression_imperfect_CSI_K_0_LS} for LS channel estimation, one can find $E_d$ so that a target probability of outage in energy transfer is maintained. One can infer how the required value of $M$ scales with the path loss $\beta$ or with the distance between the BS and the WSN, for a given $P_o$. The loss due to estimation errors can be quantified using the analysis in this section. One can also evaluate the role played by the LoS component, i.e., the Rician-$K$ factor on the outage probability using~\eqref{eq:outage_prob_perfect_CSI_K_nonzero} for perfect CSI and using~\eqref{eq:outage_prob_imperfect_CSI_LS_K_nonzero} for LS channel estimation.

\begin{table*}[t]
\caption{Probability of outage in energy transfer for different scenarios with power adaptation}
\centering
\label{table:outage_probability_with_adaptation}
\begin{tabular}{l l}
  \hline
  Scenario & Probability of outage in energy transfer \\
  \hline
  Perfect CSI, $K = 0, K \neq 0$ & 0, provided  $\rho \geq \frac{E_u+E_p}{\eta}$\\
  \hline
  LS, $K\! =\! 0$ & $\sum_{l_1=1}^{M}\!\left(\frac{2}{\mu^{\prime}}\right)^{l_1-1}\! \frac{\chi_2^{3l_1-2}}{(1-\chi_2^2)^{2l_1-1}}\! \sum_{l_2=0}^{l_1-1}\! {2l_1-l_2-2 \choose l_1-1}\!
\left(\frac{1-\chi_2^2}{\chi_2^2}\right)^{l_2}\!\left(\kappa^{\prime} {l_1-1 \choose l_2}\! -\! \chi_2 {l_1 \choose l_2}\right)\!  - \! \frac{\chi_1(\kappa^{\prime}-\chi_1)}{1-\chi_1^2}$  \\
  \hline
  MMSE estimation, $K = 0$ & $\sum_{n=1}^{M}\left(\frac{2}{\mu}\right)^{n-1} \frac{\zeta_2^{3n-2}}{(1-\zeta_2^2)^{2n-1}} \sum_{c=0}^{n-1} {2n-c-2 \choose n-1}
\left(\frac{1-\zeta_2^2}{\zeta_2^2}\right)^{c}\left(\kappa {n-1 \choose c} - \zeta_2 {n \choose c}\right) - \frac{\zeta_1(\kappa-\zeta_1)}{1-\zeta_1^2}$  \\
  \hline
  LS estimation, $K \neq 0$ & No closed-form, can be evaluated numerically using~\eqref{eq:outage_prob_imperfect_CSI_LS_K_nonzero_adapt}  \\
  \hline
  MMSE estimation, $K \neq 0$ & Single integral form, can be evaluated numerically using~\eqref{eq:exact_outage_expression_imperfect_CSI_MMSE_Knon_zero_adapt}  \\

  \hline
\end{tabular}
\end{table*}

\subsubsection{With Power Adaptation}
When $E_d = \frac{\rho}{||\widehat{\vect{h}}_{\LS}||^2}$ is varied based on the LS channel estimate, the probability of outage is given by the following result.
\begin{theorem}
\label{th:outage_prob_imperfect_CSI_LS_K_nonzero_adapt}
For a Rician fading channel and with power adaptation, the probability of outage $P_o$ in energy transfer with LS channel estimate is
\begin{multline}
P_o = \mathbb{E}\left[1  - Q_{1}\left(\sqrt{\zeta (\widehat{\vect{h}}_{\LS})}, \right. \right. \\ \left. \left. \hspace{2cm} \sqrt{\!\frac{2(E_u\!\!+\!\!E_p)\!(\beta E_u \!\!+\!\! (K+1)\! N_0)}{\eta \beta \rho N_0}}||\widehat{\vect{h}}_{\LS}||\!\!\right)\!\!\right],
\label{eq:outage_prob_imperfect_CSI_LS_K_nonzero_adapt}
\end{multline}
where $\zeta (\widehat{\vect{h}}_{\LS})$ is given by~\eqref{eq:findpdf}.
\end{theorem}
\begin{IEEEproof}
The proof is given in Appendix~\ref{app_outage_prob_imperfect_CSI_LS_K_nonzero_adapt}.
\end{IEEEproof}
Again,~\eqref{eq:outage_prob_imperfect_CSI_LS_K_nonzero_adapt} cannot be simplified any further but the expectation in~\eqref{eq:outage_prob_imperfect_CSI_LS_K_nonzero_adapt} is easily evaluated numerically. A closed-form expression for the outage probability with power adaptation and for a Rayleigh fading channel is stated below. 
\begin{corollary}
\label{th:outage_prob_imperfect_CSI_K_0_adapt}
For a Rayleigh fading channel and with power adaptation, where $\rho \geq \frac{E_u+E_p}{\eta}\left(\frac{\beta E_u + N_o}{\beta E_u}\right)^2$, the probability of outage $P_o$ in energy transfer with LS channel estimation is
\begin{multline}
P_o = \sum_{l_1=1}^{M}\left(\frac{2}{\mu^{\prime}}\right)^{l_1-1} \frac{\chi_2^{3l_1-2}}{(1-\chi_2^2)^{2l_1-1}}  \sum_{l_2=0}^{l_1-1} {2l_1-l_2-2 \choose l_1-1} \\ \times
\left(\frac{1-\chi_2^2}{\chi_2^2}\right)^{l_2}\left(\kappa^{\prime} {l_1-1 \choose l_2} - \chi_2 {l_1 \choose l_2}\right)  - \frac{\chi_1(\kappa^{\prime}-\chi_1)}{1-\chi_1^2},
\label{eq:exact_outage_expression_imperfect_CSI_K_0_LS_adapt}
\end{multline}
where $\kappa^{\prime} = \frac{\beta E_u+N_o}{\beta E_u}\sqrt{\frac{E_u+E_p}{\eta \rho}}$, $\mu^{\prime} = \frac{2 (\beta E_u+N_o)}{N_o}\sqrt{\frac{E_u+E_p}{\eta \rho}}$,
$a_0 = \sqrt{\frac{2 E_u}{N_o}\frac{\beta E_u}{(\beta E_u + N_o)}}$, $b_0=\sqrt{\frac{2(\beta E_u + N_o)}{N_o}\frac{E_u+E_p}{\eta \beta \rho}}$, $p_0 = \frac{E_u}{\beta E_u + N_o}$, $u_1 = \frac{a_0^2+b_0^2}{2a_0b_0}$, $u_2 = \frac{2p_0 + a_0^2 + b_0^2}{2a_0b_0}$, $\chi_1 = u_1-\sqrt{u_1^2-1}$, and $\chi_2 = u_2-\sqrt{u_2^2-1}$.
\end{corollary}
\begin{IEEEproof}
The proof is given in Appendix~\ref{app_outage_prob_imperfect_CSI_K_0_adapt}.
\end{IEEEproof}

\subsection{Analysis with MMSE Channel Estimation}
\label{subsec: Analysis with MMSE Channel Estimate}
In this subsection, we will analyze the probability of outage with MMSE channel estimation first with fixed $E_d$ and then with $E_d$ adapted based on the estimated channel conditions.
\subsubsection{Without Power Adaptation}
With fixed $E_d$, $P_o$ for a Rician fading channel is as follows:
\begin{theorem}
\label{th:outage_prob_imperfect_CSI_K_nonzero_MMSE}
For a Rician fading channel ($K \neq 0$), the probability of outage $P_o$ in energy transfer with MMSE channel estimate and fixed $E_d$ is
\begin{multline}
P_o = 1-\frac{2 \Lambda_1 (K+1)^{\frac{M+1}{2}}}{(K \sum_{i=0}^{M-1}\alpha_i)^{\frac{M-1}{2}}} \exp(-\Lambda_1 K \sum_{i=0}^{M-1}\alpha_i) \\ \times \int_{0}^{\infty}\!\! y_0^{M}\! \exp(-\Lambda_1 (K+1) y_0^{2})  I_{M-1}\left(\! 2\! \Lambda_1\! \sqrt{K (K+1)\sum_{i=0}^{M-1}\!\alpha_i} y_0 \!\right) \\ \times Q_{1}\left(\sqrt{\Lambda_2} y_0, \sqrt{\frac{\Lambda_2 (E_u + E_p)}{\eta \beta E_d}}\right) \, dy_0,
\label{eq:exact_outage_expression_imperfect_CSI_MMSE_Knon_zero}
\end{multline}
where  $I_{M-1}(\cdot, \cdot)$ is the $(M-1)^{\text{th}}$ order modified Bessel function of the first kind~\cite[Eqn.~(4.36)]{simon_alouini_book}, $Q_1(\cdot,\cdot)$ is the first order Marcum-Q function~\cite[Eqn.~(4.33)]{simon_alouini_book}, $\Lambda_1 = \frac{\beta E_u + (K+1) N_0}{\beta E_u}$, and $\Lambda_2 = \frac{2(\beta E_u + (K+1) N_0)}{N_0}$.
\end{theorem}
\begin{IEEEproof}
The proof is given in Appendix~\ref{app_outage_prob_imperfect_CSI_K_nonzero_MMSE}.
\end{IEEEproof}

Note that~\eqref{eq:exact_outage_expression_imperfect_CSI_MMSE_Knon_zero} is in the form of a single integral in $y_0$ and probably cannot be simplified any further as the integrand involves the product of a modified Bessel function and a Marcum-Q function. It is, however, easy to evaluate numerically. An integral-free closed-form expression for the outage probability for a Rayleigh fading channel can be obtained as stated below.
\begin{corollary}
\label{th:outage_prob_imperfect_CSI_K_0_MMSE}
For a Rayleigh fading channel ($K = 0$), the probability of outage $P_o$ in energy transfer for a fixed $E_d$ and with MMSE channel estimation is
\begin{multline}
P_o = 1-\frac{\beta E_u}{\beta E_u + N_o} \exp\left(-\frac{E_u+E_p}{\eta \beta E_d}\right) \\ \hspace{1cm} \times \sum_{k=0}^{M-1} \epsilon_{k} \left(\frac{N_o}{\beta E_u + N_o}\right)^{k} L_{k}\left(-\frac{E_u(E_u+E_p)}{\eta E_d N_o}\right),
\label{eq:exact_outage_expression_imperfect_CSI_MMSE}
\end{multline}
where $L_{k}(\cdot)$ is the $k^{\text{th}}$ Laguerre polynomial and $\epsilon_{k}$ is given by~\eqref{eq:lag_coeff}.
\end{corollary}
\begin{IEEEproof}
The proof is given in  Appendix~\ref{app_outage_prob_imperfect_CSI_K_0_MMSE}.
\end{IEEEproof}

Note that the expressions for the probability of outage in energy transfer is the same for both the MMSE and LS estimators under i.i.d. Rayleigh fading and  for fixed $E_d$. This is because under i.i.d. Rayleigh fading, the LS and the MMSE estimates differ only in terms of the scaling factor. Since, we normalize the beamforming vector, this scaling has no impact on the end performance. However, under spatial correlation or even for uncorrelated Rician fading, MMSE gives a better outage performance compared to LS estimator, the difference being significant particularly at low SNRs as shown in Section~\ref{Correlated_channels} and in Figures~\ref{fig:Impact of estimation error_Rician_rev2} and~\ref{fig:Impact of estimation error under spatial correlation} in Section~\ref{sec:Numerical Results}.

\begin{table}
\caption{Estimates of Path Loss and the Corresponding BS-WSN Separation from~\cite{Visser_2013_ieee}}
\centering
\label{table:pathloss_estimate}
\begin{tabular}{l l}
  \hline
  Path loss ($\beta$) & BS-WSN distance \\
  \hline
  60 dB &7.8 m  \\
   \hline
   55 dB &4.1 m  \\
   \hline
    50 dB &2.2 m  \\
    \hline
     45 dB &1.1 m  \\
     \hline
\end{tabular}
\end{table}

\subsubsection{With Power Adaptation}
If, however, $E_d = \frac{\rho}{||\hhat_{\MMSE}||^2}$ is adapted based on the MMSE estimate, the probability of outage in energy transfer for a Rician fading channel is as follows:
\begin{theorem}
\label{th:outage_prob_imperfect_CSI_K_nonzero_MMSE_adapt}
For a Rician fading channel ($K \neq 0$) and with power adaptation, the probability of outage $P_o$ in energy transfer with MMSE channel estimate is
\begin{multline}
P_o = 1-\frac{2 \Lambda_1 (K+1)^{\frac{M+1}{2}}}{(K \sum_{i=0}^{M-1}\alpha_i)^{\frac{M-1}{2}}} \exp(-\Lambda_1 K \sum_{i=0}^{M-1}\alpha_i) \\ \times \!\! \int_{0}^{\infty}\!\!\!\! y_0^{M}\! \exp(-\Lambda_1 (K+1) y_0^{2})  I_{M-1}\left(\!2 \!\Lambda_1 \!\sqrt{K (K+1) \! \sum_{i=0}^{M-1} \! \alpha_i}  y_0 \! \right) \\ \times Q_{1}\left(\sqrt{\Lambda_2} y_0, \sqrt{\frac{\Lambda_2 (E_u + E_p)}{\eta \rho}}y_0\right) \, dy_0.
\label{eq:exact_outage_expression_imperfect_CSI_MMSE_Knon_zero_adapt}
\end{multline}
\end{theorem}
\begin{IEEEproof}
The proof is given in Appendix~\ref{app_outage_prob_imperfect_CSI_K_nonzero_MMSE_adapt}.
\end{IEEEproof}

Note that~\eqref{eq:exact_outage_expression_imperfect_CSI_MMSE_Knon_zero_adapt} is in the form of a single integral in $y_0$ and probably cannot be simplified any further as the integrand involves the product of a modified Bessel function and a Marcum-Q function. It is, however, easily evaluated numerically. The outage probability for a Rayleigh fading channel with power control is given by the following result.
\begin{corollary}
\label{th:outage_prob_imperfect_CSI_K_0_MMSE_adapt}
For a Rayleigh fading channel ($K  = 0$) and with power adaptation where $\rho \geq \frac{E_u+E_p}{\eta}$, the probability of outage $P_o$ in energy transfer with MMSE channel estimate is
\begin{multline}
P_o = \sum_{n=1}^{M}\left(\frac{2}{\mu}\right)^{n-1} \frac{\zeta_2^{3n-2}}{(1-\zeta_2^2)^{2n-1}} \sum_{c=0}^{n-1} {2n-c-2 \choose n-1}
\\ \times \left(\frac{1-\zeta_2^2}{\zeta_2^2}\right)^{c}\left(\kappa {n-1 \choose c} - \zeta_2 {n \choose c}\right) - \frac{\zeta_1(\kappa-\zeta_1)}{1-\zeta_1^2},
\label{eq:exact_outage_expression_imperfect_CSI_MMSE_adapt}
\end{multline}
where $\kappa = \sqrt{\frac{E_u+E_p}{\eta \rho}}$, $\mu = \frac{2 \beta E_u}{N_o}\sqrt{\frac{E_u+E_p}{\eta \rho}}$,
$a = \sqrt{\frac{2(\beta E_u + N_o)}{\beta N_o}}$, $b=\sqrt{\frac{2(\beta E_u + N_o)}{N_o}\frac{E_u+E_p}{\eta \beta \rho}}$, $p = \frac{\beta E_u + N_o}{\beta^2 E_u}$, $v_1 = \frac{a^2+b^2}{2ab}$, $v_2 = \frac{2p + a^2 + b^2}{2ab}$, $\zeta_1 = v_1-\sqrt{v_1^2-1}$, and $\zeta_2 = v_2-\sqrt{v_2^2-1}$.
\end{corollary}
\begin{IEEEproof}
The proof is given in  Appendix~\ref{app_outage_prob_imperfect_CSI_K_0_MMSE_adapt}.
\end{IEEEproof}

\subsection{Extensions to Non-Identical and Correlated Channels}
\label{subsec:Extensions to Non-Identical and Correlated Channels}
\subsubsection{Independent and Non-Identically Distributed Rayleigh Fading Channels}
For analytical tractability, we focus on the scenario where the $M$ wireless links from the BS to the WSN see independent but non-identically distributed (i.n.i.d.) Rayleigh fading and derive the outage probability with perfect CSI. We do not discuss the case with LS or MMSE channel estimation as they are analytically intractable and do not provide any additional insights.
\begin{theorem}
\label{th:outage_prob_imperfect_CSI_K_nonzero_MMSE_adapt}
For i.n.i.d Rayleigh fading channels without power adaptation and with perfect CSI, the probability of outage $P_o$ in energy transfer is
\begin{equation}
P_o = \sum_{j=1}^{M} \frac{\beta_j^{M-1}\left(1-\exp\left(-\frac{E_u+E_p}{\eta \beta_j E_d}\right)\right)}{\prod_{k=1,k\neq j}^{M}(\beta_j-\beta_k)}.
\label{eq:exact_outage_expression_perfect_CSI_inid}
\end{equation}
\end{theorem}
\begin{IEEEproof}
The proof involves finding the CDF of $||\vect{h}||^2$, where each $h_i \sim \mathcal{CN}\left(0, \beta_i\right)$ and which can be obtained using the result in~\cite[Eqn.~(3)]{Ho_2006_commLet}.
\end{IEEEproof}

\subsubsection{Correlated Channels}
\label{Correlated_channels}
To analyze the effect of spatial correlation on outage probability, we generate the channel covariance matrix $\mtx{R}$ using the exponential correlation matrix model from~\cite{Loyka_2001_commLet}. For this model, the $(i, j)$th element of $\mtx{R}$ is given by
\begin{equation}
[\mtx{R}]_{i,j} = \begin{cases} r^{j-i}, &  i \leq j, \\
\left(r^{i-j}\right)^{\ast} & i>j. \end{cases}
\end{equation}
This model basically represents a uniform linear array where the  correlation coefficient between adjacent antennas is given by $|r|$, for $0 \leq |r| \leq 1$.

The outage probability analysis developed in this section gives us insights about the feasibility of WET using multi-antenna arrays.


\section{Numerical Results}
\label{sec:Numerical Results}

In this section, we present numerical results to quantify the potential of using massive antenna arrays for WET using the two phase communication scheme in Figure~\ref{fig:system_model_two_phases}. Unless mentioned otherwise, we take $E_u = 10^{-8}$~J (e.g., 100 $\mu$W during 100 $\mu$s), $E_d = 10^{-3}$~J (e.g., 1 W during 1 ms), $E_p = 10^{-7}$~J (e.g., 1 mW during 100 $\mu$s), $\eta = 0.5$,\footnote{The rectenna efficiency depends on the RF input power because the rectifier consists of a diode whose impedance varies non-linearly as a function of the input power level. The efficiency also depends on the rectenna design, the power management unit, the frequency of operation, and the load connected to the sensor node~\cite{Popovic_2013_procIEEE, Popovic_2014_ieeeMicrowave}. As illustrated in Figure~7 in~\cite{Popovic_2013_procIEEE}, for a specific rectenna design, if the RF power input is above a certain threshold, the efficiency does not vary much, whereas below this threshold (which depends on the technology used), the efficiency reduces. For the numerical example given in Figure~7 of~\cite{Popovic_2013_procIEEE}, it can be observed that the rectenna efficiency reaches almost $45$\% at an input power level of $1$~mW and increases marginally to $50$\% when the input power level is around $6$~mW. In our numerical examples, we use an efficiency figure of  $\eta=0.5$ and  make sure that the input RF power is sufficient to justify this value. For example, with $E_d = 10^{-3}$~J, $\beta = -50$~dB, and $M = 30$, the received energy  is about $30 \times 10^{-8}$~J. Over a $100$~$\mu$s duration, this corresponds to a received power of $3$~mW. So, an appropriate justifiable choice of the time duration for the receive signal would ensure that the input power is in the order of a mW. As the path loss increases or $E_d$ reduces, more antennas can be deployed to ensure that the received RF power remains the same.
As mentioned above, if the RF power received at the sensor is lower than a certain threshold (1 mW for the cited rectenna) then the energy harvesting efficiency reduces and a value less than $0.5$ (depending on the input power) will have to be considered~\cite{Popovic_2013_procIEEE}. Note that an intermediate power management unit is needed to impedance match the rectenna to the energy storage unit in the sensor node. The design of the associated power management unit  is discussed in~\cite{Paing_2008_ieeePower, Dolgov_2010_ieeeCircuits, Paing_2011_ieeePower}. Received power levels on the order of a mW are envisioned in  applications like charging of hearing aids, wireless keyboards and mouses~\cite{Xiao_2015_CommSurvey}. As another example, the commercially available equipment Powercast (TX91501) harvests about 4 mW with a 3 W transmitter at a range of 1 meter at 915 MHz carrier frequency. Another potential application of our work could be in millimeter-wave cellular networks  where received power on the order of a mW is considered reasonable~\cite{Wang_2015_arxiV}. }
and $N_0 = k_B T 10^{F/10} = 10^{-20}$~J, where $k_B = 1.38 \times 10^{-23}$~J/K, $T = 300$~K, and the receiver noise figure is $F = 7$~dB. Also, we consider a uniform linear array\footnote{Each antenna in the array is omnidirectional, only the array as a whole can form a beam and not each antenna on its own.} for which $\theta_i(\phi) = 2 \pi d i \cos (\phi)$, $i = 1, \ldots, M-1$. We take $\phi = \pi/3$ and $d = \frac{\lambda}{2} = 0.06$~m, where $\lambda$ is the wavelength at a frequency of $2.45$~GHz. We also detail below estimation of typical product of path loss and energy harvesting efficiency from experimental results in~\cite{Visser_2013_ieee}, that we will use in our numerical examples.

{\em Estimation of path loss and energy harvesting efficiency from experiments in~\cite{Visser_2013_ieee}:}
From~\cite{Visser_2013_ieee}, when a $4$~W transmitter connected to a vertically polarized fan beam array antenna (with gain $G_T = 9 $~dB) is employed, the DC power harvested as a function of the distance from the transmitter in a LoS situation is plotted in Figure~3 of~\cite{Visser_2013_ieee}. The carrier frequency ($f_c$) used in the experiment in~~\cite{Visser_2013_ieee} is $2.45$~GHz. Also, this experiment was carried out in an office corridor environment. From this plot, the product $\eta \beta$ of the energy harvesting efficiency and the path loss can be estimated as follows:
\begin{equation}
\eta \beta = \frac{P_{\text{DC}}}{P_T G_T G_R}
\label{eq:estimate eta beta}
\end{equation}
where $P_{\text{DC}}$ is the DC power harvested by the sensor, $P_T$ is the transmit power, $G_T$, and $G_R$ are the gains of the transmit and receive antennas respectively. We assume that $G_R = 0$~dB.
Figure~3 given in~\cite{Visser_2013_ieee} shows that at a distance of $1.1$~m from the transmitter, the DC power harvested is about $0.5$~mW. Therefore, from~\eqref{eq:estimate eta beta}, $\eta \beta = \frac{P_{\text{dc}}}{P_T G_T G_R} = 1.58 \times 10^{-5} = -48~\rm{dB}$.
Assuming an energy harvesting efficiency $\eta = 0.5$, this gives us a path loss of $\beta = -45$~dB at a distance of $1.1$~m. Similarly, different estimates of the path loss and the corresponding distances between the BS and the WSN can be obtained as listed in Table~\ref{table:pathloss_estimate}. We vary $\beta$ around a nominal value of $-50$~dB in our examples.

Our numerical results are based on analytical expressions developed in the paper and
since the Monte Carlo simulations overlap with these and do not provide any additional information, they are not shown.\footnote{Results with LS estimator also overlap with that of the MMSE estimator for typical energy levels required to enable WET.}

\begin{figure}
\centering
\includegraphics[width=0.9\linewidth]{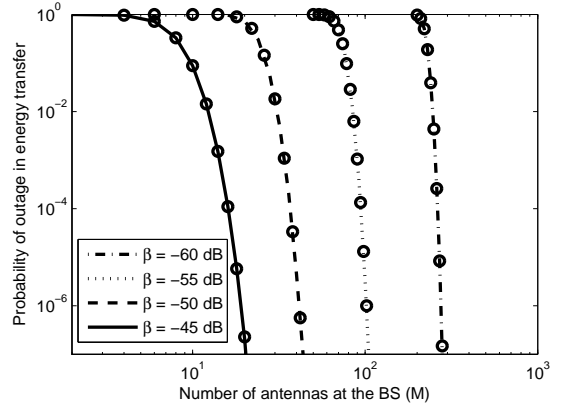}
\caption{MMSE/LS channel estimation: Impact of $\beta$ and $M$ on $P_o$ ($E_d = 10^{-3}$~J, $E_u = 10^{-8}$~J, $E_p = 10^{-7}$~J, $N_0 = 10^{-20}$~J, $K = 2$, $\alpha_i = 1$, for all $i = 0, \ldots, M-1$, and $\eta = 0.5$). The corresponding perfect CSI results are shown using `$\circ$'.}
\label{fig:Imperfect CSI impact of M and beta}
\end{figure}

Figure~\ref{fig:Imperfect CSI impact of M and beta} plots $P_o$ as a function of $M$ for different values of $\beta$ and for $K = 2$. We observe that by deploying more antennas at the BS, a larger path loss (larger distance between the BS and the WSN) can be tolerated while keeping the outage probability fixed. For example, by going from about $20$ antennas to $100$ antennas at the BS, an outage probability of $10^{-6}$ can be maintained even if the path loss increases from $45$~dB to $55$~dB. Also, for $E_u = 10^{-8}$~J and $E_d = 10^{-3}$~J, the difference in performance among MMSE, LS, and perfect CSI is negligibly small.

\begin{figure}
\centering
\includegraphics[width=0.9\linewidth]{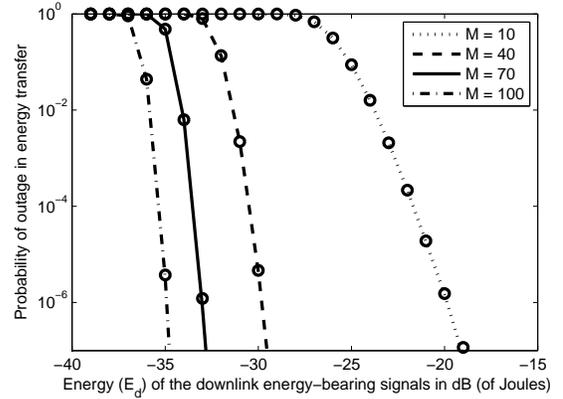}
\caption{MMSE/LS channel estimation: Impact of $E_d$ and $M$ on $P_o$ ($\beta \!=\! -50$~dB, $E_u \!=\! 10^{-8}$~J, $E_p \!=\! 10^{-7}$~J, $N_0 \!=\! 10^{-20}$~J, $K = 2$, $\alpha_i = 1$, for all $i = 0, \ldots, M-1$, and $\eta \!=\! 0.5$). The corresponding perfect CSI results are shown using `$\circ$'.}
\label{fig:outage vs Ed imperfect CSI}
\end{figure}

Figure~\ref{fig:outage vs Ed imperfect CSI} plots $P_o$ as a function of $E_d$ for different values of $M$ and for $K=2$. It can be observed that as $E_d$ increases, the outage probability decreases. Moreover, as more antennas are deployed at the BS, a lower $E_d$ is required to keep the outage probability at the same value. For example, by going from about $10$ to $40$ antennas at the BS, $E_d$ can be reduced by $8$~dB, while keeping the outage probability fixed at $10^{-6}$. Thus, the array gain obtained by deploying multiple antennas at the BS results in huge savings of radiated energy. One can also see that the difference in performance among MMSE, LS, and perfect CSI is negligibly small at practical operating points.

\begin{figure}
\centering
\includegraphics[width=0.9\linewidth]{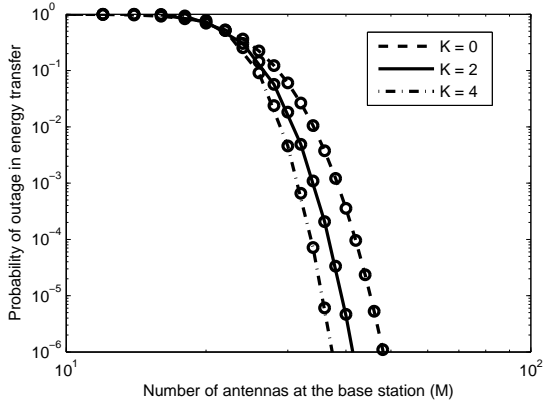}
\caption{MMSE/LS channel estimation: Impact of $K$ and $M$ on $P_o$ ($E_u = 10^{-8}$~J, $E_p = 10^{-7}$~J, $E_d = 10^{-3}$~J, $\beta = -50$~dB, $N_0 = 10^{-20}$~J, $\alpha_i = 1$, for all $i = 0, \ldots, M-1$, and $\eta = 0.5$). The corresponding perfect CSI results are shown using `$\circ$'.}
\label{fig:outage vs M imperfect CSI_nonzero_K}
\end{figure}

Figure~\ref{fig:outage vs M imperfect CSI_nonzero_K} plots $P_o$ as a function of $M$ for three different values of the Rician-$K$ factor, namely, $K = 0$, $K = 2$, and $K = 4$ and for both perfect and imperfect CSI obtained again using MMSE/LS channel estimation. It can be observed that as $K$ increases, the channel becomes more deterministic and the outage probability improves with perfect or imperfect CSI. In other words, a strong line-of-sight component in the channel helps in lowering the outage probability. Also, for the energy levels $E_u = 10^{-8}$~J and $E_d = 10^{-3}$~J, the difference in performance among MMSE, LS, and perfect CSI is negligibly small. Thus, at these energy levels, one can as well use LS channel estimation instead of an MMSE estimator that assumes a priori knowledge of the angle of arrival, basically $\vect{h}_d$, the Rician-$K$ factor, the channel and noise distributions without degrading the outage probability relative to the perfect CSI scenario.

\begin{figure}
\centering
\includegraphics[width=0.9\linewidth]{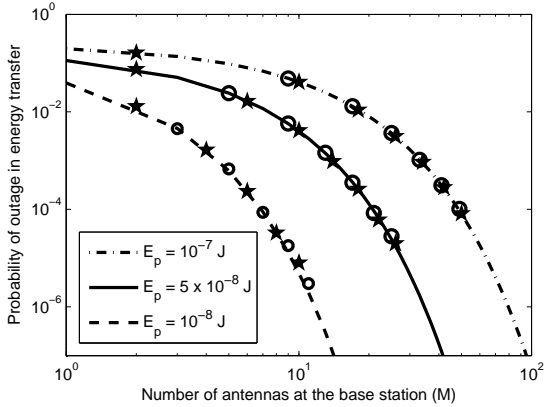}
\caption{I.n.i.d channels: Impact of $E_p$ and $M$ on $P_o$ ($E_u = 10^{-8}$~J, $E_d = 10^{-3}$~J, $N_0 = 10^{-20}$~J, K = 0, $\mtx{\Lambda}_h = \text{diag}(\beta_1, \ldots, \beta_M)$, and $\eta = 0.5$). Perfect CSI results from~\eqref{eq:exact_outage_expression_perfect_CSI_inid} are shown using different line types, corresponding MMSE results using `$\circ$' and LS using `$\star$'.}
\label{fig:impact_Ep}
\end{figure}

Figure~\ref{fig:impact_Ep} plots $P_o$ as a function of $M$ for different values of the processing energy $E_p$ and for i.n.i.d. channels. It can be observed that by deploying more antennas at the BS, the WSN gets higher amount of processing energy to perform its main tasks at a given target outage probability. Thus, multiple antennas at the BS can help transfer more energy to the energy-constrained WSNs.

\begin{figure}
\centering \includegraphics[width=0.9\linewidth]{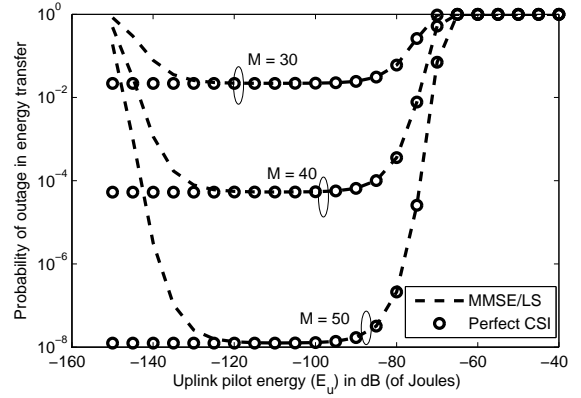}
\caption{Impact of uplink pilot energy on $P_o$ ($E_d = 10^{-3}$~J,  $E_p = 10^{-7}$~J, $N_0 = 10^{-20}$~J, $\beta = -50$~dB, $K = 0$, $r=0$, and $\eta = 0.5$).}
\label{fig:Impact of uplink_pilot_power_rev2}
\end{figure}
As shown in Figure~\ref{fig:Impact of uplink_pilot_power_rev2}, with MMSE/LS channel estimation, a too high uplink pilot energy would lead to an increase in the outage probability as a larger portion of the harvested energy is spent on training. A too low uplink pilot energy would again increase the outage probability due to increased channel estimation errors. Note that for i.i.d. Rayleigh fading channels, MMSE and LS estimation give the same performance irrespective of $E_u$. With perfect CSI, for smaller $E_u$'s the outage performance essentially remains the same, since $E_u$ is much smaller than $E_p$.

\begin{figure}
\centering \includegraphics[width=0.9\linewidth]{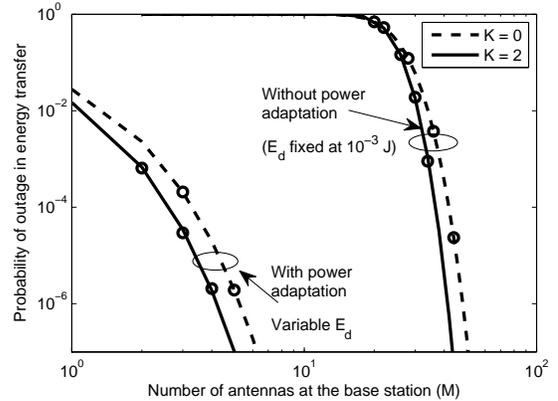}
\caption{Impact of power control on $P_o$ ($N_0 = 10^{-20}$~J, $\beta = -50$~dB, $E_p = 10^{-7}$~J, $E_u = 10^{-8}$~J, $\rho = 2.204 \times 10^{-7}$~J, $\alpha_i = 1$, for all $i = 0, \ldots, M-1$, and $\eta = 0.5$. MMSE results are shown using linetypes and LS results using `$\circ$').}
\vspace{-0.6cm}
\label{fig:impact_power control}
\end{figure}

Figure~\ref{fig:impact_power control} plots $P_o$ as a function of $M$ for two different values of $K$ with and without power adaptation for both LS and MMSE estimation. As expected, power adaptation at the BS helps to improve the outage probability. For the case when $E_d$ is adapted based on the channel conditions, while with perfect CSI, the outage probability is zero irrespective of $M$, with MMSE/LS channel estimation it is non-zero but decays very quickly to zero as $M$ increases for an appropriately chosen value of $\rho$ for both $K=0$ and $K=2$. Note that we choose $\rho$ to be greater than or equal to $(E_u+E_p)/\eta$ otherwise, the outage probability is $1$ even with perfect CSI and as stated in Theorem~2.

\begin{figure}
\centering
\includegraphics[width=0.9\linewidth]{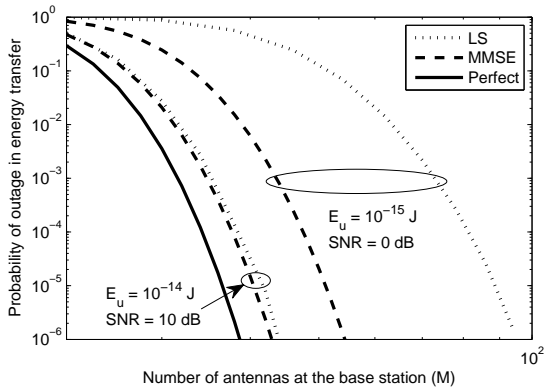}
\caption{Impact of estimation error on $P_o$ ($E_d = 10^{-3}$~J,  $E_p = 10^{-7}$~J, $N_0 = 10^{-20}$~J, $\beta = -50$~dB, $K = 2$, $\alpha_i = 1$, for all $i = 0, \ldots, M-1$, and $\eta = 0.5$).}
\vspace{-0.6cm}
\label{fig:Impact of estimation error_Rician_rev2}
\end{figure}

Figure~\ref{fig:Impact of estimation error_Rician_rev2} plots $P_o$ as a function of $M$ for perfect CSI, LS, and MMSE channel estimation and for two different values of the uplink pilot energy $E_u$, basically the low SNR regime for uncorrelated Rician fading. Please note that the outage performance with perfect CSI is the same for both $E_u = 10^{-14}$~J or $E_u = 10^{-15}$~J since $E_u$ is much smaller than $E_p$. While MMSE performs marginally better than LS estimation at $E_u = 10^{-14}$~J, the gap in performance is significantly higher at $E_u = 10^{-15}$~J.

\begin{figure}
\centering \includegraphics[width=0.9\linewidth]{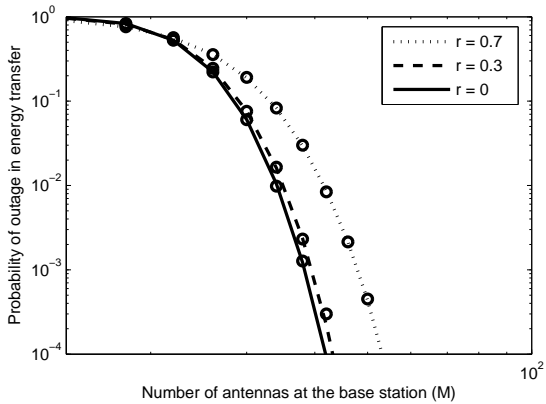}
\caption{MMSE/LS channel estimation: Impact of spatial correlation $r$ on $P_o$ ($E_d = 10^{-3}$~J, $E_u = 10^{-8}$~J, $E_p = 10^{-7}$~J, $N_0 = 10^{-20}$~J, $\beta = -50$~dB, $K = 0$, and $\eta = 0.5$). The corresponding perfect CSI results are shown using `$\circ$'.}
\vspace{-0.6cm}
\label{fig:Impact_of_spatial_correlation}
\end{figure}
\begin{figure}
\centering \includegraphics[width=0.9\linewidth]{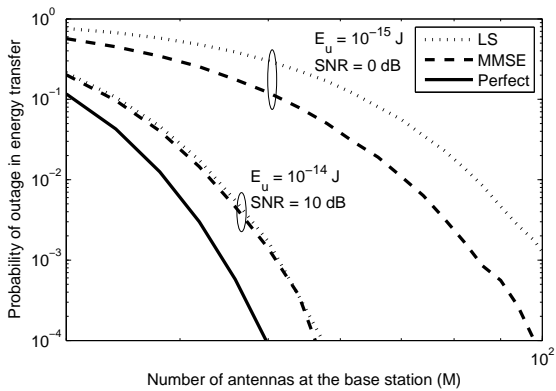}
\caption{Impact of estimation error under spatial correlation on $P_o$ ($E_d = 10^{-3}$~J,  $E_p = 10^{-7}$~J, $N_0 = 10^{-20}$~J, $\beta = -50$~dB, $K = 0$, $r = 0.7$ , and $\eta = 0.5$).}
\vspace{-0.6cm}
\label{fig:Impact of estimation error under spatial correlation}
\end{figure}

Figure~\ref{fig:Impact_of_spatial_correlation} plots $P_o$ as a function of $M$ for different values of $r$. As expected, with an increase in antenna correlation, the probability of outage in energy transfer deteriorates. In other words, by deploying more antennas at the base station, a higher antenna correlation factor can be tolerated while maintaining a given target outage probability. Note that for the chosen parameters, MMSE and LS channel estimation perform as well as perfect CSI for all values of $r$.

Figure~\ref{fig:Impact of estimation error under spatial correlation} plots $P_o$ as a function of $M$ for perfect CSI, LS, and MMSE channel estimation and for two different values of the uplink pilot energy $E_u$ for spatially correlated Rayleigh fading channels. From this figure, it is clear that MMSE outperforms LS channel estimation under spatial correlation and at low signal-to-noise ratios. While at $E_u = 10^{-14}$~J, MMSE performs marginally better than LS estimation, the gap is significant for $E_u = 10^{-15}$~J. Note that with perfect CSI, the outage performance is the same for both $E_u = 10^{-14}$~J or $E_u = 10^{-15}$~J, because $E_u$ is very small compared to $E_p$.

\section{Conclusions}
\label{sec: Conclusions}

We investigated the feasibility of using multiple antennas at the transmitter for WET. Specifically, we derived expressions for the outage probability when the BS uses an array of antennas to focus and transfer energy to a WSN and where the channel from the array to the WSN is estimated using pilots sent by the WSN. This is done both with perfect CSI and with LS or MMSE channel estimates and for Rayleigh fading (without dominant components) and Rician fading (with dominant components). We proved that by adding more antennas at the BS, we can extend the range for WET while maintaining a given target outage probability. We further observed that a lower downlink energy is required to get the same performance due to huge array gains obtained by multi-antenna beamforming.

We observed that for the typical energy levels that are used in WET, the difference in outage performance among LS, MMSE, and perfect CSI is negligibly small. Further, we show that a strong LoS component between the BS and the WSN helps improve the outage probability. We also show that by deploying more antennas at the BS, a larger energy can be transferred to the WSN for it to be able to perform its main tasks. While with perfect CSI, outage can be completely eliminated by power adaptation based on the channel conditions. With power adaptation based on LS or MMSE estimate, it can be considerably reduced.

\appendix

\subsection{Proof of Lemma~\ref{Lem_LS}}
\label{app_conditional_stats_Psi_LSestimate}
Conditioned on $\widehat{\vect{h}}_{\LS}$, $\Psi_{\LS}$ is a complex Gaussian RV with mean
\begin{equation}
 \expect{\Psi_{\LS}|\widehat{\vect{h}}_{\LS}}  =  \frac{\widehat{\vect{h}}_{\LS}^{\dag}}{||\widehat{\vect{h}}_{\LS}||}\expect{\vect{h}|\widehat{\vect{h}}_{\LS}}.
 \label{eq:basic_LS_proof}
\end{equation}

Using standard results on conditional Gaussian RVs~\cite{kay_book_vol1}, it can be shown that
\begin{multline}
\expect{\vect{h}|\widehat{\vect{h}}_{\LS}}  = \vect{\mu} + \frac{\beta}{K+1}\mathbf{I}_M \left(\frac{\beta E_u + (K+1)N_0}{E_u (K+1)}\mathbf{I}_M\right)^{-1} \\ \times \left(\widehat{\vect{h}}_{\LS} - \vect{\mu}\right).
\end{multline}
Substituting $\expect{\vect{h}|\widehat{\vect{h}}_{\LS}}$ in~\eqref{eq:basic_LS_proof} and simplifying yields~\eqref{cond_mean_LS}. Also, the conditional variance is
\begin{equation}
\var{\Psi_{\LS}|\widehat{\vect{h}}_{\LS}} = \frac{\widehat{\vect{h}}_{\LS}^{\dag} \rm{cov}\left(\vect{h}|\widehat{\vect{h}}_{\LS}\right) \widehat{\vect{h}}_{\LS}}{||\widehat{\vect{h}}_{\LS}||^2}.
\end{equation}

Again, using standard results on conditional Gaussian RVs~\cite{kay_book_vol1}, it can be shown that $\mathrm{cov}\left(\vect{h}|\widehat{\vect{h}}_{\LS}\right) = \frac{\beta N_0}{\beta E_u + (K+1) N_0} \mathbf{I}_M$.
Therefore, the conditional variance of $\Psi_{\LS}$ simplifies to~\eqref{cond_var_LS}.

\subsection{Proof of Lemma~\ref{Lem_MMSE}}
\label{app_conditional_stats_Psi_MMSEestimate}
Conditioned on $\widehat{\vect{h}}_{\MMSE}$, $\Psi_{\MMSE}$ is a complex Gaussian RV with mean
\begin{equation}
 \expect{\Psi_{\MMSE}|\widehat{\vect{h}}_{\MMSE}}  =  \frac{\widehat{\vect{h}}_{\MMSE}^{\dag}}{||\widehat{\vect{h}}_{\MMSE}||}\expect{\vect{h}|\widehat{\vect{h}}_{\MMSE}}.
\end{equation}
Using standard results on conditional Gaussian RVs~\cite{kay_book_vol1}, it is easy to show that
%
\begin{multline}
\expect{\vect{h}|\widehat{\vect{h}}_{\MMSE}} = \left(\vect{\mu} + \frac{\beta^2 E_u}{(\beta E_u+(K+1)N_0)(K+1)}\mathbf{I}_M \right. \\ \left. \times  \left(\frac{\beta^2 E_u}{(\beta E_u + (K+1)N_0)(K+1)}\mathbf{I}_M\right)^{-1} \left(\widehat{\vect{h}}_{\MMSE} - \vect{\mu}\right)\right)
\end{multline}
Upon simplification, this yields $ \expect{\Psi_{\MMSE}|\widehat{\vect{h}}_{\MMSE}} = ||\widehat{\vect{h}}_{\MMSE}||$.

And conditional variance is given by
\begin{equation}
\var{\Psi_{\MMSE}|\widehat{\vect{h}}_{\MMSE}} = \frac{\widehat{\vect{h}}_{\MMSE}^{\dag} \rm{cov}(\vect{h}|\widehat{\vect{h}}_{\MMSE}) \widehat{\vect{h}}_{\MMSE}}{||\widehat{\vect{h}}_{\MMSE}||^2}.
\end{equation}

Again using standard results on conditional Gaussian RVs~\cite{kay_book_vol1}, it can be shown that
\begin{equation}
\mathrm{cov}(\vect{h}|\widehat{\vect{h}}_{\MMSE}) = \frac{\beta N_0}{\beta E_u + (K+1) N_0} \mathbf{I}_M.
\end{equation}
Therefore, the conditional variance of $\Psi_{\MMSE}$ simplifies to~\eqref{eq:cond_var}.

\subsection{Proof of Theorem~\ref{th:outage_prob_perfect_CSI_K_nonzero}}
\label{app_outage_prob_perfect_CSI_K_nonzero}
With perfect CSI, $\Psi = \Psi_{\LS} = \Psi_{\MMSE} =  \frac{\hhat^{\dag} \vect{h}}{||\hhat||} = ||\vect{h}||$,
where $\hhat = \vect{h}$ can be either the LS or the MMSE channel estimate.
Therefore,
\begin{align}
\nonumber
P_o & = \Pr(\eta E_d |\Psi|^2 \leq E_u + E_p) \\
\nonumber
& = \Pr\left(\frac{2  (K+1)}{\beta} ||\vect{h}||^2 \leq  \frac{2 (K+1) (E_u+E_p)}{\eta \beta E_d}\right) \\
\label{eq:exact_outage_perfect_CSI_K_nonzero}
& = 1-Q_{M}\left(\sqrt{2 K \sum_{i=0}^{M-1}\alpha_i}, \sqrt{\frac{2(K+1) (E_u+E_p)}{\eta \beta E_d}}\right).
\end{align}
Note that~\eqref{eq:exact_outage_perfect_CSI_K_nonzero} follows from the fact that $\frac{2 (K+1)}{\beta} ||\vect{h}||^2$
is a non-central chi-square distributed RV with $2 M$ degrees of freedom and non-centrality parameter $2 K \sum_{i=0}^{M-1}\alpha_i$.

\subsection{Proof of Theorem~\ref{th:outage_prob_perfect_CSI_K_nonzero_adapt}}
\label{app_outage_prob_perfect_CSI_K_nonzero_adapt}
With power adaptation, $E_d = \frac{\rho}{||\vect{h}||^2}$.
Therefore, the probability of outage in energy transfer is
\begin{equation}
P_o  = \Pr(\eta E_d ||\vect{h}||^2 \leq E_u + E_p) = \Pr\left(\rho \leq \frac{E_u + E_p}{\eta}\right).
\end{equation}
Since the expression is deterministic, $P_o$ is one if $\rho \leq \frac{E_u+E_p}{\eta}$. If, on the other hand, $\rho \geq \frac{E_u+E_p}{\eta}$, then the energy harvested $E_h = \eta E_d ||\vect{h}||^2= \eta \rho \geq E_u+E_p$ and there will never be an outage.

\
\subsection{Proof of Theorem~\ref{th:outage_prob_imperfect_CSI_LS_K_nonzero}}
\label{app_outage_prob_imperfect_CSI_LS_K_nonzero}

With LS channel estimation,
\begin{equation}
\label{eq:initial_outage_LS_Rician}
P_o  = \expect{\Pr\left(|\Psi_{\LS}|^2 \leq \frac{E_u+E_p}{\eta E_d} \middle | \widehat{\vect{h}}_{\LS}\right)}.
\end{equation}
Let
$\widetilde{\Psi}_{\LS} = \frac{\Psi_{\LS}}{\sqrt{\frac{\beta N_0}{2(\beta E_u + (K+1) N_0)}}}$.
Therefore, $P_o$ in~\eqref{eq:initial_outage_LS_Rician} can be written as
\begin{equation}
P_o  = \expect{\Pr \left(|\psit_{\LS}|^2 \leq \frac{2(E_u+E_p)(\beta E_u + (K+1) N_0)}{\eta \beta E_d N_0}\middle | \widehat{\vect{h}}_{\LS}\right)}.
\label{fin_LS}
\end{equation}

Using Lemma~\ref{Lem_LS}, it can be shown that given $\hhat_{\LS}$, $\mathrm{Re}(\psit_{\LS})$ and $\mathrm{Im}(\psit_{\LS})$ are independent Gaussian RVs with conditional statistics
$\expect{\mathrm{Re}(\psit_{\LS})|\hhat_{\LS}} = \frac{\sqrt{2}\left(\beta E_u ||\hhat_{\LS}||^2 + \mathrm{Re}\left(\hhat_{\LS}^{\dag}\vect{\mu}\right)(K+1)N_0\right)}{\sqrt{\beta N_0 \left(\beta E_u + (K+1) N_0\right)}||\hhat_{\LS}||}$, $\expect{\mathrm{Im}(\psit_{\LS})|\hhat_{\LS}} =
\frac{\sqrt{2 N_0} (K+1)}{\sqrt{\beta(\beta E_u + (K+1) N_0)}}\frac{\mathrm{Im}\left(\widehat{\vect{h}}_{\LS}^{\dag}\vect{\mu}\right)}{||\widehat{\vect{h}}_{\LS}||}$ and
$\var{\mathrm{Re}(\psit_{\LS})|\hhat_{\LS}} = \var{\mathrm{Im}(\psit_{\LS})|\hhat_{\LS}} = 1$.

Thus, given $\hhat_{\LS}$, $|\psit_{\LS}|^2$ is a non-central chi-square distributed RV with $2$ degrees of freedom and non-centrality parameter given by
\begin{multline}
\zeta (\widehat{\vect{h}}_{\LS}) =  \frac{2\left(\beta E_u ||\hhat_{\LS}||^{2} + \mathrm{Re}\left(\hhat_{\LS}^{\dag}\vect{\mu}\right) (K+1) N_0 \right)^2}{\beta N_0(\beta E_u + (K+1) N_0)||\hhat_{\LS}||^{2}}  \\ + \frac{2 N_0 (K+1)^{2}}{\beta(\beta E_u + (K+1) N_0)}\left(\frac{\mathrm{Im}\left(\widehat{\vect{h}}_{\LS}^{\dag}\vect{\mu}\right)}{||\widehat{\vect{h}}_{\LS}||}\right)^{2}.
\label{eq:proof_findpdf_proof}
\end{multline}
Substituting the cumulative distribution function (CDF) of $|\psit_{\LS}|^2$ given $\widehat{\vect{h}}_{\LS}$ in~\eqref{fin_LS} yields~\eqref{eq:outage_prob_imperfect_CSI_LS_K_nonzero}.


\subsection{Proof of Corollary~\ref{th:outage_prob_imperfect_CSI_K_0}}
\label{app_outage_prob_imperfect_CSI_K_0}
For a Rayleigh fading channel, by substituting $K = 0$ in~\eqref{eq:outage_prob_imperfect_CSI_LS_K_nonzero}, we get
\begin{multline}
P_o  = \mathbb{E}\left[1 - Q_{1}\left(\sqrt{\frac{2 E_u}{N_0}\frac{\beta E_u}{\beta E_u+N_0}} ||\hhat_{\LS}|| , \right. \right. \\ \left. \left. \hspace{2cm}\sqrt{ \frac{2(\beta E_u + N_0)(E_u+E_p)}{\eta \beta E_d N_0}}\right)\right],
\label{eq:intermediate_outage_result}
\end{multline}
where $Q_{1}(\cdot,\cdot)$ is the first order Marcum-Q function~\cite[Eqn~(4.33)]{simon_alouini_book}.

To compute~\eqref{eq:intermediate_outage_result}, we need to find the distribution of $Y = ||\hhat_{\LS}|| = \sqrt{|\hat{h}_{\LS_{1}}|^{2} + \cdots + |\hat{h}_{\LS_{M}}|^{2}}$.
Note that for $K = 0$, $\hat{h}_{\LS_{i}} \sim \mathcal{CN}\left(0, \frac{\beta E_u + N_0}{E_u}\right)$. This implies that $\frac{2 E_u}{\beta E_u+ N_0} Y^{2}$
 is a chi-square distributed RV with $2 M$ degrees of freedom since it is the sum of the squares of $2 M$ independent standard normal RVs.  Therefore, the RV $Z = Y^{2}$ has the PDF given by
\begin{equation}
f_{Z}(z) \!=\! \left(\frac{E_u}{\beta E_u+N_0}\right)^{M}\! \frac{z^{M-1}}{(M-1)!}\! \exp\left(\frac{-E_u z}{\beta E_u+N_0}\right), ~z \geq 0.
\end{equation}
By transformation of RVs, it can be shown that $Y = \sqrt{Z} = ||\hhat_{\LS}||$ has the PDF given by
\begin{equation}
f_{Y}(y) \!=\! 2 \!\left(\!\frac{E_u}{\beta E_u+N_0}\!\right)^{M}\!\!\! \frac{y^{2 M-1}}{(M-1)!}\! \exp\left(\!\frac{-E_u y^2}{\beta E_u+N_0}\!\right), ~y \geq 0.
\label{eq:pdf_Y}
\end{equation}
Substituting the PDF of $Y$ from~\eqref{eq:pdf_Y} in~\eqref{eq:intermediate_outage_result}, we get
\begin{multline}
 P_o = 1- \frac{2\left(\frac{E_u}{\beta E_u+N_0}\right)^{M}}{(M-1)!} \int_{0}^{\infty} y^{2 M - 1} \exp\left(\frac{-E_u y^2}{\beta E_u+N_0}\right) \\ \times Q_{1}\left(\!\!\sqrt{\frac{2 E_u}{N_0}\frac{\beta E_u}{\beta E_u+N_0}} y, \sqrt{\frac{2(\beta E_u + N_0)(E_u+E_p)}{\eta \beta E_d N_0}}\!\!\right) \, dy.
\label{eq:outage_final_Rayleigh_LS}
\end{multline}
Note the variable $y$ in just one of the arguments of the Marcum-Q function.
Using the identity in~\cite[Eqn.~(9)]{Nuttall_1975_ieeeIt},~\eqref{eq:outage_final_Rayleigh_LS} can be simplified to yield~\eqref{eq:exact_outage_expression_imperfect_CSI_K_0_LS}.

\subsection{Proof of Theorem~\ref{th:outage_prob_imperfect_CSI_LS_K_nonzero_adapt}}
\label{app_outage_prob_imperfect_CSI_LS_K_nonzero_adapt}

With LS channel estimation and with power adaptation,
\begin{equation}
\label{eq:initial_outage_LS_Rician_adapt}
P_o  = \expect{\Pr\left(\frac{|\Psi_{\LS}|^2}{||\widehat{\vect{h}}_{\LS}||^2} \leq \frac{E_u+E_p}{\eta \rho} \middle | \widehat{\vect{h}}_{\LS}\right)}.
\end{equation}
Let $\widetilde{\Psi}_{\LS} = \frac{\Psi_{\LS}}{\sqrt{\frac{\beta N_0}{2(\beta E_u + (K+1) N_0)}}}$.
Therefore, $P_o$ in~\eqref{eq:initial_outage_LS_Rician_adapt} can be written as
\begin{multline}
P_o  \\ = \mathbb{E}\left[\!\Pr \left(\!\frac{|\psit_{\LS}|^2}{||\widehat{\vect{h}}_{\LS}||^2} \!\leq \!\frac{2\!(E_u\!+\!E_p)\!(\beta E_u \!+\! (K+1) N_0)}{\eta \beta \rho N_0}\middle | \widehat{\vect{h}}_{\LS}\!\right)\!\right].
\label{fin_LS_adapt}
\end{multline}

From Appendix~\ref{app_outage_prob_imperfect_CSI_LS_K_nonzero}, we know that given $\hhat_{\LS}$, $|\psit_{\LS}|^2$ is a non-central chi-square distributed RV with $2$ degrees of freedom and non-centrality parameter given by~\eqref{eq:proof_findpdf_proof}.
Substituting the CDF of $|\psit_{\LS}|^2$ given $\widehat{\vect{h}}_{\LS}$ in~\eqref{fin_LS_adapt} yields~\eqref{eq:outage_prob_imperfect_CSI_LS_K_nonzero_adapt}.

\subsection{Proof of Corollary~\ref{th:outage_prob_imperfect_CSI_K_0_adapt}}
\label{app_outage_prob_imperfect_CSI_K_0_adapt}
For a Rayleigh fading channel, by substituting $K = 0$ in~\eqref{eq:outage_prob_imperfect_CSI_LS_K_nonzero_adapt}, we get
\begin{multline}
P_o  = \mathbb{E}\left[1  - Q_{1}\left(\sqrt{\frac{2 E_u}{N_0}\frac{\beta E_u}{\beta E_u+N_0}} ||\hhat_{\LS}||, \right. \right. \\ \left. \left. \hspace{2cm}\sqrt{ \frac{2(\beta E_u+N_0)(E_u+E_p)}{\eta \rho \beta N_0}}||\hhat_{\LS}||\right)\right].
\label{eq:intermediate_outage_result_adapt}
\end{multline}

To compute~\eqref{eq:intermediate_outage_result_adapt}, we need the distribution of $Y = ||\hhat_{\LS}|| = \sqrt{|\hat{h}_{\LS_{1}}|^{2} + \cdots + |\hat{h}_{\LS_{M}}|^{2}}$ that we have already evaluated in~\eqref{eq:pdf_Y}.
Substituting the PDF of $Y$ from~\eqref{eq:pdf_Y} in~\eqref{eq:intermediate_outage_result_adapt}, we get
\begin{multline}
P_o  = 1- \frac{2\left(\frac{E_u}{\beta E_u+N_0}\right)^{M}}{(M-1)!} \int_{0}^{\infty} y^{2 M - 1} \exp\left(\frac{-E_u y^2}{\beta E_u+N_0} \right) \\ \times Q_{1}\left(\sqrt{\frac{2 E_u}{N_0}\frac{\beta E_u}{\beta E_u+N_0}} y, \sqrt{ \frac{2(\beta E_u+N_0)(E_u+E_p)}{\eta \rho \beta N_0}}y\right) \, dy.
\label{last_Gaur_LS}
\end{multline}
Note the variable $y$ in both the arguments of the Marcum-Q function.
If $\rho \geq \frac{E_u+E_p}{\eta}\left(\frac{\beta E_u + N_o}{\beta E_u}\right)^2$,~\eqref{last_Gaur_LS} can be simplified using the identity in~\cite[Eqn.~(25)]{Gaur_2003_ieeeVt} to obtain~\eqref{eq:exact_outage_expression_imperfect_CSI_K_0_LS_adapt}.


\subsection{Proof of Theorem~\ref{th:outage_prob_imperfect_CSI_K_nonzero_MMSE}}
\label{app_outage_prob_imperfect_CSI_K_nonzero_MMSE}
With MMSE channel estimation and for a Rician fading channel,
\begin{equation}
P_o = \expect{\Pr\left(|\Psi_{\MMSE}|^2 \leq \frac{E_u+E_p}{\eta E_d} \middle | \widehat{\vect{h}}_{\MMSE}\right)}.
\label{eq:initial_outage_MMSE_rician}
\end{equation}
Let
$\widetilde{\Psi}_{\MMSE} = \frac{\Psi_{\MMSE}}{\sqrt{\frac{\beta N_0}{2(\beta E_u +(K+1) N_0)}}}$.
Therefore,~\eqref{eq:initial_outage_MMSE_rician} reduces to
\begin{multline}
 P_o\! \\ =\! \expect{\!\Pr \!\left(\!|\psit_{\MMSE}|^2 \!\leq \! \frac{2(\!E_u\!+\!E_p\!)\! \left(\!\beta E_u \!+\! (K+1) N_0\!\right)}{\eta \beta E_d N_0} \middle | \widehat{\vect{h}}_{\MMSE}\!\right)\!}. \\
\label{eq:initial_outage_result_MMSE_K_nonzero}
\end{multline}

Given $\hhat_{\MMSE}$, $\mathrm{Re}(\psit_{\MMSE})$ and $\mathrm{Im}(\psit_{\MMSE})$ are independent Gaussian RVs. Using Lemma~\ref{Lem_MMSE}, it can be shown that
$\expect{\mathrm{Re}(\psit_{\MMSE})|\hhat_{\MMSE}} = \sqrt{2\left(\frac{\beta E_u + (K+1) N_0}{\beta N_0}\right)} ||\hhat_{\MMSE}||$, $\expect{\mathrm{Im}(\psit_{\MMSE})|\hhat_{\MMSE}} = 0$, and the conditional variances are given by
$\var{\mathrm{Re}(\psit_{\MMSE})|\hhat_{\MMSE}} = \var{\mathrm{Im}(\psit_{\MMSE})|\hhat_{\MMSE}} = 1$.
Thus, given $\hhat_{\MMSE}$, $|\psit_{\MMSE}|^2$ is a non-central chi-square distributed RV with $2$ degrees of freedom and non-centrality parameter $2\left(\frac{\beta E_u  + (K+1) N_0}{\beta N_0}\right) ||\hhat_{\MMSE}||^2$.
Therefore,~\eqref{eq:initial_outage_result_MMSE_K_nonzero} reduces to
\begin{equation}
P_o = \expect{1-Q_{1}\left(\sqrt{\Lambda_0} ||\hhat_{\MMSE}||, \sqrt{\frac{\Lambda_0 (E_u+E_p)}{\eta E_d}}\right)},
\label{eq:intermediate}
\end{equation}
where $\Lambda_0 = 2 \left(\frac{\beta E_u + (K+1) N_0}{\beta N_0}\right)$.

To compute~\eqref{eq:intermediate}, we need to find the distribution of $Y_0 = ||\hhat_{\MMSE}|| = \sqrt{|\hat{h}_{\MMSE_{1}}|^{2} + \cdots + |\hat{h}_{\MMSE_{M}}|^{2}}$. It can be shown that $\frac{2(K+1) \left(\beta E_u + (K+1) N_0\right)}{\beta^2 E_u} Y_0^{2} $
is a non-central chi-square distributed RV with $2 M$ degrees of freedom and non-centrality parameter $\frac{2 K \sum_{i=0}^{M-1}\alpha_i\left(\beta E_u + (K+1) N_0\right)}{\beta E_u}$. Therefore, the RV $Z_0 = Y_0^{2}$ has the PDF
\begin{multline}
f_{Z_0}(z_0) = \frac{(K+1)^{\frac{M+1}{2}}}{(K \beta \sum_{i=0}^{M-1}\alpha_i)^{\frac{M-1}{2}}}  \Lambda z_0^{\frac{M-1}{2}} \\ \times  \exp\left(- \Lambda \left((K+1)z_0 + K\beta \sum_{i=0}^{M-1}\alpha_i\right)\right) \\ \times I_{M-1}\left(2 \Lambda \sqrt{\beta K (K+1)\sum_{i=0}^{M-1}\alpha_i z_0}\right), ~~~z_0 \geq 0,
\end{multline}
where $\Lambda = \frac{\beta E_u + (K+1) N_0}{\beta^2 E_u}$ and $I_{M-1}(\cdot)$ is the modified Bessel function of the $(M-1)^{\text{th}}$ order and first kind.

By transformation of RVs, it can be shown that $Y_0 = \sqrt{Z_0} = ||\hhat_{\MMSE}||$ has the PDF
\begin{multline}
f_{Y_0}(y_0) = \frac{2 \Lambda (K+1)^{\frac{M+1}{2}}}{(K \beta \sum \limits_{i=0}^{M-1}\alpha_i)^{\frac{M-1}{2}}} \exp(-\Lambda K \beta \sum_{i=0}^{M-1}\alpha_i) y_0^{M} \\ \times \exp(-\Lambda (K+1) y_0^{2})  I_{M-1}\left(2 \Lambda \sqrt{\beta K (K+1)\sum_{i=0}^{M-1}\alpha_i} y_0\right).
\label{eq:pdf_Y_MMSE_nonzeroK}
\end{multline}
Substituting the PDF of $Y_0$ from~\eqref{eq:pdf_Y_MMSE_nonzeroK} in~\eqref{eq:intermediate} and a simple change of variables yields~\eqref{eq:exact_outage_expression_imperfect_CSI_MMSE_Knon_zero}.

\subsection{Proof of Corollary~\ref{th:outage_prob_imperfect_CSI_K_0_MMSE}}
\label{app_outage_prob_imperfect_CSI_K_0_MMSE}
For a Rayleigh fading channel, by substituting $K = 0$ in~\eqref{eq:intermediate}, we get
\begin{multline}
P_o = \mathbb{E}\left[1-Q_{1}\left(\sqrt{\frac{2\left(\beta E_u+ N_0\right)}{\beta N_0}} ||\hhat_{\MMSE}||, \right. \right. \\ \left. \left. \hspace{2cm} \sqrt{\frac{2(\beta E_u + N_0)(E_u+E_p)}{\eta \beta E_d N_0}}\right)\right].
\label{eq:intermediate_outage_result_MMSE}
\end{multline}
To compute~\eqref{eq:intermediate_outage_result_MMSE}, we need to find the distribution of $Y_1 = ||\hhat_{\MMSE}|| = \sqrt{|\hat{h}_{\MMSE_{1}}|^{2} + \cdots + |\hat{h}_{\MMSE_{M}}|^{2}}$.
Note that for $K = 0$, $\hat{h}_{\MMSE_{i}} \sim \mathcal{CN}\left(0, \frac{\beta^2 E_u}{\beta E_u + N_0}\right)$. This implies that $\frac{2 \left(\beta E_u + N_0\right)}{\beta^2 E_u} Y_1^{2}$
is a chi-square distributed RV with $2 M$ degrees of freedom since it is the sum of the squares of $2 M$ independent standard normal RVs.  Therefore, the RV $Z_1 = Y_1^{2}$ has the PDF (for $z_1 \geq 0$)
\begin{equation}
f_{Z_1}(z_1) \!=\! \left(\!\frac{\beta E_u + N_0}{\beta^2 E_u}\!\right)^{M}\!\! \frac{z_1^{M-1}}{(M-1)!}\! \exp\left(\frac{-(\beta E_u + N_0)z_1}{\beta^2 E_u}\right).
\end{equation}
By transformation of RVs, it can be shown that $Y_1 = \sqrt{Z_1} = ||\hhat_{\MMSE}||$ has the PDF (for $y_1 \geq 0$)
\begin{equation}
f_{Y_1}(y_1) \!=\! 2 \left(\!\frac{\beta E_u + N_o}{\beta^2 E_u}\!\right)^{M}\! \frac{y_1^{2 M-1}}{(M-1)!}\! \exp\left(\frac{-(\beta E_u + N_o)y_1^2}{\beta^2 E_u}\right).
\label{eq:pdf_Y_MMSE}
\end{equation}
Substituting the PDF of $Y_1$ from~\eqref{eq:pdf_Y_MMSE} in~\eqref{eq:intermediate_outage_result_MMSE}, we get
\begin{multline}
P_o = 1- \frac{2\left(\frac{\beta E_u + N_o}{\beta^2 E_u}\right)^{M}}{(M-1)!} \int_{0}^{\infty}\!\! y_1^{2 M - 1} \exp\left(\frac{-(\beta E_u + N_o)y_1^2}{\beta^2 E_u}\right) \\ \times Q_{1}\left(\!\!\sqrt{2\left(\frac{\beta E_u+N_0}{\beta N_0}\right)} y_1, \sqrt{2 \left(\frac{\beta E_u+N_0}{N_0}\right)\!\!\frac{(E_u+E_p)}{\eta \beta E_d}}\!\!\right) \, dy_1.
\label{eq:outage_MMSE_rayleigh}
\end{multline}
Note the variable $y_1$ in just one of the arguments of the Marcum-Q function.
Using the identity in~\cite[Eqn.~(9)]{Nuttall_1975_ieeeIt},~\eqref{eq:outage_MMSE_rayleigh} can be simplified to yield~\eqref{eq:exact_outage_expression_imperfect_CSI_MMSE}.

\subsection{Proof of Theorem~\ref{th:outage_prob_imperfect_CSI_K_nonzero_MMSE_adapt}}
\label{app_outage_prob_imperfect_CSI_K_nonzero_MMSE_adapt}
With MMSE channel estimation for a Rician fading channel and with power adaptation,
\begin{equation}
P_o = \expect{\Pr\left(\frac{|\Psi_{\MMSE}|^2}{||\hhat_{\MMSE}||^2} \leq \frac{E_u+E_p}{\eta \rho} \middle | \widehat{\vect{h}}_{\MMSE}\right)}.
\label{eq:initial_outage_MMSE_rician_adapt}
\end{equation}
Let
$\widetilde{\Psi}_{\MMSE} = \frac{\Psi_{\MMSE}}{\sqrt{\frac{\beta N_0}{2(\beta E_u +(K+1) N_0)}}}$
and rewrite~\eqref{eq:initial_outage_MMSE_rician_adapt} as
\begin{multline}
 P_o\!=\! \mathbb{E}\left[\!\Pr \!\left(\!\frac{|\psit_{\MMSE}|^2}{||\hhat_{\MMSE}||^2} \!\leq \!  \right. \right.\\ \left. \left. \hspace{2cm}\frac{2 (E_u+E_p) \left(\beta E_u + (K+1) N_0\right)}{\eta \beta \rho N_0} \middle | \widehat{\vect{h}}_{\MMSE}\!\right)\!\right]. \\
\label{eq:initial_outage_result_MMSE_K_nonzero_adapt}
\end{multline}

From Appendix~\ref{app_outage_prob_imperfect_CSI_K_nonzero_MMSE}, we know that given $\hhat_{\MMSE}$, $|\psit_{\MMSE}|^2$ is a non-central chi-square distributed RV with $2$ degrees of freedom and non-centrality parameter $2\left(\frac{\beta E_u  + (K+1) N_0}{\beta N_0}\right) ||\hhat_{\MMSE}||^2$.
Therefore,~\eqref{eq:initial_outage_result_MMSE_K_nonzero_adapt} reduces to
\begin{equation}
P_o = \expect{1-Q_{1}\left(\sqrt{\Lambda_0} ||\hhat_{\MMSE}||, \sqrt{\frac{\Lambda_0 (E_u+E_p)}{\eta \rho}} ||\hhat_{\MMSE}||\right)}.
\label{eq:intermediate_adapt}
\end{equation}

Substituting the PDF of $||\hhat_{\MMSE}||$ from~\eqref{eq:pdf_Y_MMSE_nonzeroK} in~\eqref{eq:intermediate_adapt} and a simple change of variables yields~\eqref{eq:exact_outage_expression_imperfect_CSI_MMSE_Knon_zero_adapt}.

\subsection{Proof of Corollary~\ref{th:outage_prob_imperfect_CSI_K_0_MMSE_adapt}}
\label{app_outage_prob_imperfect_CSI_K_0_MMSE_adapt}
For a Rayleigh fading channel, by substituting $K = 0$ in~\eqref{eq:intermediate_adapt}, we get
\begin{multline}
P_o = \mathbb{E}\left[1-Q_{1}\left(\sqrt{2\left(\frac{\beta E_u+N_0}{\beta N_0}\right)} ||\hhat_{\MMSE}||, \right. \right.\\ \left. \left. \hspace{2cm}\sqrt{2 \left(\frac{\beta E_u+N_0}{\beta N_0}\right)\frac{E_u+E_p}{\eta \rho}}||\hhat_{\MMSE}||\right)\right].
\label{eq:intermediate_outage_result_MMSE_adapt}
\end{multline}

Substituting the PDF of $||\hhat_{\MMSE}||$ from~\eqref{eq:pdf_Y_MMSE} in~\eqref{eq:intermediate_outage_result_MMSE_adapt}, we get
\begin{multline}
P_o \!=\! 1- \frac{2\left(\frac{\beta E_u+N_o}{\beta^2 E_u}\right)^{M}}{(M-1)!} \int_{0}^{\infty}\!\! y_1^{2 M - 1} \exp\left(\frac{-(\beta E_u + N_o) y_1^2}{\beta^2 E_u} \right) \\ \times Q_{1}\left(\sqrt{\frac{2(\beta E_u+N_0)}{\beta N_0}} y_1, \sqrt{\frac{2(\beta E_u+N_0)}{\beta N_0}\frac{E_u+E_p}{\eta\rho}} y_1\!\!\right) dy_1.
\label{last_Gaur}
\end{multline}
Note the variable $y_1$ in both the arguments of the Marcum-Q function.
For $\rho \geq \frac{E_u+E_p}{\eta}$,~\eqref{last_Gaur} can be simplified using the identity in~\cite[Eqn.~(25)]{Gaur_2003_ieeeVt} to obtain~\eqref{eq:exact_outage_expression_imperfect_CSI_MMSE_adapt}.


\end{document}